\def\ts     {\thinspace}
\def\kms    {\ifmmode{{\rm \ts km\ts s}^{-1}}\else{\ts km\ts s$^{-1}$}\fi}
\def\msol     {\ifmmode{{\rm M}_{\odot}}\else{M$_{\odot}$}\fi}
\def\etal   {{\rm et\ts al.}}
\def\aco  {\ifmmode{^{12}{\rm CO}(J\!=\!1\! \to \!0)}\else{$^{12}{\rm CO}(J\!=\!1\! \to \!0)$}\fi}
\def\bco  {\ifmmode{^{12}{\rm CO}(J\!=\!2\! \to \!1)}\else{$^{12}{\rm CO}(J\!=\!2\! \to \!1)$}\fi}
\def\m    {\ifmmode{\mu {\rm m}}\else{$\mu$m}\fi}
\def\cco  {\ifmmode{^{13}{\rm CO}(J\!=\!1\! \to \!0)}\else{$^{13}{\rm CO}(J\!=\!1\! \to \!0)$}\fi}
\def\dco  {\ifmmode{^{13}{\rm CO}(J\!=\!2\! \to \!1)}\else{$^{13}{\rm CO}(J\!=\!2\! \to \!1)$}\fi}
\def\eco  {\ifmmode{{\rm C}^{18}{\rm O}(J\!=\!1\! \to \!0)}\else{${\rm C}^{18}{\rm O}(J\!=\!1\! \to \!0)$}\fi}
\def\nh   {\ifmmode{N(\hi)}\else{$N$(\hi)}\fi}
\def\hun    {\ifmmode{I_{100}}\else{$I_{100}$}\fi}
\def\sex    {\ifmmode{I_{60}}\else{$I_{60}$}\fi}
\def\hh     {\ifmmode{{\rm H}_2}\else{H$_2$}\fi}
\def\nhh     {\ifmmode{N({\rm H}_2)}\else{$N$(H$_2$)}\fi}
\def\zwco   {\ifmmode{^{12}{\rm CO}}\else{$^{12}{\rm CO}$}\fi}
\def\nzwco   {\ifmmode{N(^{12}{\rm CO})}\else{$N(^{12}{\rm CO})$}\fi}
\def\wzwco   {\ifmmode{W(^{12}{\rm CO})}\else{$W(^{12}{\rm CO})$}\fi}
\def\drco   {\ifmmode{^{13}{\rm CO}}\else{$^{13}{\rm CO}$}\fi}
\def\ndrco   {\ifmmode{N(^{13}{\rm CO})}\else{$N(^{13}{\rm CO})$}\fi}
\def\wdrco   {\ifmmode{W(^{13}{\rm CO})}\else{$W(^{13}{\rm CO})$}\fi}
\def\tex    {\ifmmode{T_{ex}({\rm CO})}\else{$T_{ex}({\rm CO})$}\fi}
\def\ha     {\ifmmode{{\rm H}\alpha}\else{${\rm H}\alpha$}\fi}
\def\amm     {\ifmmode{{\rm NH}_{3}}\else{${\rm NH}_{3}$}\fi}
\def\xco     {\ifmmode{X_{\rm CO}}\else{$X_{\rm CO}$}\fi}
\def\tkin     {\ifmmode{T_{\rm kin}}\else{$T_{\rm kin}$}\fi}
\documentclass[a4paper,structabstract]{aa}
\usepackage{txfonts}
\usepackage{graphicx}
\usepackage{epsfig}
\usepackage{multirow}
\usepackage{ulem}
\begin{document}
   \title{Dense gas in nearby galaxies}

\subtitle{XVII. The distribution of ammonia in NGC\,253, Maffei\,2
and IC\,342}

   \author{ M. Lebr\'on\inst{1,2}
   \and
      J. G. Mangum\inst{3}
      R. Mauersberger\inst{4,5}
          \and
          C. Henkel\inst{2,6}
          \and
          A.B. Peck\inst{4,3}
          \and
      K.M. Menten\inst{2}
          \and
          A. Tarchi\inst{7}
          \and
          A. Wei{\ss}\inst{2}
      }

\offprints{J.~G.~Mangum, email: jmangum@nrao.edu}
   \institute{
             Department of Physical Sciences, University of Puerto
             Rico, P.O. Box 23323, San Juan 00931-3323, Puerto Rico
              \and
             Max-Planck-Institut f\"ur Radioastronomie,
             Auf dem H\"ugel 69,
             D-53121 Bonn, Germany
             \and
             National Radio Astronomy Observatory,
             520 Edgemont Road,
             Charlottesville, VA  22903
             \and
             Joint ALMA Observatory, Av. Alonso de C\'ordova 3107, Vitacura,
             Santiago, Chile 
             \and
             Instituto de Radioastronom\'{\i}a Milim\'etrica,
             Avda. Divina Pastora 7, Local\,20, E-18012 Granada, Spain
             \and
             Astronomy Department, Faculty of Science, King Abdulaziz
             University, P.~O.~Box 80203, Jeddah, Saudi Arabia 
             \and
             INAF-Osservatorio
             Astronomico di Cagliari, Loc. Poggio dei Pini,
             Strada 54, I-09012 Capoterra (CA), Italy
             }
             \date{Received September 15, 2222; accepted March 16, 3222}

   \abstract
   {The central few 100 pc of galaxies often contain large
   amounts of molecular gas. The chemical and physical properties of
   these extragalactic star formation regions differ from
   those in galactic disks, but are poorly constrained.}
   {This study aims to develop a better knowledge of the spatial
   distribution and kinetic temperature of the dense neutral gas
   associated with the 
   nuclear regions of three prototypical spiral galaxies, NGC\,253,
   IC\,342, and Maffei\,2.}
   {VLA CnD and D configuration measurements have been made of three ammonia
   (NH$_3$) inversion transitions.} 
   {The $(J,K)=(1,1)$ and (2,2) transitions of NH$_3$ were imaged toward
   IC\,342 and Maffei\,2. The (3,3) transition was imaged toward
   NGC\,253.  The entire flux obtained from
   single-antenna measurements is recovered for all three
   galaxies observed.  Derived lower limits to the kinetic temperatures 
   determined for the giant molecular clouds in the centers of these
   galaxies are between 25 and 50\,K.  There is good agreement
   between the distributions of NH$_3$ and other H$_2$ 
   tracers, such as rare CO isotopologues or HCN, suggesting that NH$_3$ is
   representative of the distribution of dense gas.  The
   ``Western Peak'' in IC\,342 is seen in the (6,6) line but not
   in lower transitions, suggesting maser emission in the (6,6)
   transition.}
   {}
   \maketitle
%

\section{Introduction}
\label{Intro}
A large fraction ($\sim 20\%$ at $z<0.2$, Brinchmann et al.
\cite{brinchmann04}) of all stars, and as a consequence many of the
massive stars in our Universe, are being born in the central regions
of starburst galaxies. The reservoirs for such starbursts are large
concentrations of dense molecular gas, which in many cases are formed
by the interaction of two or more galaxies, which triggers the infall
of large quantities of material toward the central few hundred parsecs in
these systems (Combes \cite{combes2005}). The typical sizes of the
clouds formed in such environments are a few 10s of parsecs.
Observations to-date suggest that the physical properties of the
molecular clouds in these extragalactic environments are different
from Giant Molecular Clouds (GMCs) found in the disks of the Milky Way
and other normal galaxies. The chemistry of such extragalactic
circumnuclear clouds is comparable in richness to the 
Orion or Sgr\,B2 star forming regions, as a line survey at 2\,mm
toward NGC\,253 suggests (Mart\'in et al. \cite{martin06b}). The spatial
densities within these clouds should be $\ga 10^4\,\rm cm^{-3}$,
i.e. higher than in disk clouds, in order to counteract the tidal
forces induced by a high stellar density and a supermassive central
engine (G\"usten \cite{guesten89}). Such densities have indeed been
confirmed by a number of molecular multilevel studies
(e.g. Mauersberger et al. \cite{mau95}, Wei{\ss} et
al. \cite{weiss2007}, Costagliola et
al. \cite{cost2010}, Aladro et al. \cite{aladro2011}). As in the
circumnuclear region of our own Galaxy 
(Oka et al. \cite{oka2005}), these dense clouds may be embedded in a
less dense ($\sim 100\,\rm cm^{-3}$) and warm ($\sim$250\,K) molecular
component.

Ammonia (NH$_3$) is the classical tracer of the kinetic temperature
(\tkin) within the dense neutral interstellar medium (ISM; see Ho \&
Townes \cite{ho83}). Unlike most other molecules, where the transition
excitation depends both on the density of molecular hydrogen, $n(\rm
H_2)$, and kinetic temperature, \tkin, the relative transition
intensities of the inversion transitions of this symmetric 
top molecule depend mainly on $T_{\rm kin}$ for a large range of
densities. In addition, NH$_3$ has numerous inversion transitions at centimeter
wavelengths covering a large range of energies; it therefore probes a
wide range of temperatures. While brightness 
temperatures of main isotopic CO transitions are indicators of $T_{\rm
kin}$ only for nearby dark clouds, where beam filling factors are
close to unity, NH$_3$ multilevel studies can be used to determine
kinetic temperatures even in cases where there is sub-structure
smaller than the size of the beam. This is particularly useful for
extragalactic sources. 

The first, and for a long time the only,
extragalactic ammonia study was limited to IC\,342, a nearby face-on
galaxy with particularly narrow spectral lines (Martin \&
Ho \cite{martin86}). Thanks to improved  sensitivity and spectral
baseline stability, several extragalactic NH$_3$ multilevel studies 
have been published during the past decade for the nearby galaxies NGC\,253,
M\,51, M\,82, and Maffei\,2 (Henkel et al. \cite{hen00}; Takano et
al. \cite{takano00}; Wei{\ss} et al. \cite{weiss01a}; Mauersberger et
al. \cite{mau03}; Ott et al. 
\cite{ott05b}).  Kinetic temperatures have been estimated toward 
the central regions of these galaxies on spatial scales of several
hundred pc.  Takano et al. (\cite{Takano05a}) also presented Very
Large Array (VLA) images of the NH$_3$ (1,1), (2,2) and (3,3) emission
from NGC\,253, and Montero-Casta\~no et al. 
(\cite{montero2006}) presented a VLA (6,6) image of IC\,342.
NH$_3$ absorption lines have been reported toward Arp\,220 by
Takano et al. \cite{takano2005b} and, at intermediate redshift,
toward the main gravitational lenses of the radio sources
B0218+357 and PKS\,1830--211 (Henkel et al. \cite{henkel05}, 
\cite{henkel08}).

In this paper, interferometric observations of the $(J,K)=(1,1)$ and
$(2,2)$ ammonia inversion transitions are presented for the central
regions of the nearby galaxies IC\,342 and Maffei\,2, and of the
$(3,3)$ transition toward NGC\,253. The linear resolutions of these
new NH$_3$ measurements are $\sim 50$\,pc.  The primary goals are
to determine the spatial distribution and to constrain the kinetic
temperature of the NH$_3$ emitting gas, and to relate these to the
dominant physical and chemical parameters of the studied regions.

\section{Observations and data reduction}
\label{ObsDataRed}
The $(J,K)$ = (1,1) and (2, 2) inversion transitions of NH$_3$,
(rest frequencies: 23.6944955 GHz and 23.7226333 GHz, Lovas
\cite{lovas04}), were observed toward IC\,342 and Maffei\,2 on
22--27 October 2001 using the D configuration of the Very Large
Array (VLA) of the NRAO\footnote{The National Radio Astronomy
Observatory is a facility of the National Science Foundation
operated under cooperative agreement by Associated Universities,
Inc.}. The beam sizes are $3\farcs 8 \times 3\farcs 4$ at a P.A. $-85^\circ$
for IC\,342 and $3\farcs 6 \times 3\farcs 2$ at a P.A. $-89^\circ$ for
Maffei\,2. We used a bandpass of 25 MHz centered at heliocentric
velocities of $+40$ km s$^{-1}$ for IC\,342. The heliocentric velocity of
Maffei\,2 is $-$40 \kms, but due to limitations in the LO tuning at
the VLA for bandpasses of 25~MHz or more, we set the central
velocity to $-$35 \kms. The total bandwidth for both sources was
divided into 15 spectral channels, each 1.5625 MHz wide ($\sim$19.7 km
s$^{-1}$ at the observing frequencies), plus a continuum channel
containing the central 75\% of the total band. For the amplitude
calibration, we used 0137$+$331, with an adopted flux density of 1.05~Jy.
The phase calibrators were 0304$+$683 with a flux density of 0.43
Jy, and 0244$+$624 with a flux density of 1.6 Jy at 1.3\,cm.

The NH$_3$(3,3) transition (rest frequency: 23.8701292 GHz, Lovas
\cite{lovas04}) was observed toward NGC\,253 on 29--30 September
2001 using the VLA DnC configuration. This configuration was
required because of the low elevation of this source. The resulting
beam size was $2\farcs 9 \times 2\farcs 2$ at a P.A. 58$^\circ$. The
observations used a bandwidth of 50 MHz centered at a heliocentric
velocity of 9\,km\,s$^{-1}$ and 16 spectral channels, each of them
covering $\sim$ 40\,km\,s$^{-1}$.  Since the systemic
velocity of NGC\,253 is at $\sim 230$\,km\,s$^{-1}$, the transition is
at the edge of the observed band. From a comparison with CO spectra
(e.g. Mauersberger et al. \cite{mau03}) we estimate, however, that 
$\la 5\%$ of the flux is outside the observed spectral
bandpass. Absolute flux density calibration was obtained by observing
0137+331 (1.05\,Jy). Phase and bandpass calibration were performed on
0120$-$270.

The data were edited and calibrated following the standard VLA
procedures and using the Astronomical Image Processing System (AIPS)
developed by the NRAO. Observations from different days were
combined in the UV plane. We synthesized, cleaned, and naturally weighted maps of
the observed regions. For each source, a continuum map was produced by
averaging line-free channels.

In order to convert flux density per beam, $S$, to synthesized beam
brightness temperature, $T_{\rm b}$, we have used the relation

\begin{equation}
S{\rm (Jy/beam)}/T_{\rm b}=8.18\,10^{-7}\,\theta_{\rm
a}({\rm asec})\theta_{\rm b}({\rm asec})(\nu ({\rm GHz}))^2
\label{eq:jyperk}
\end{equation}

\noindent{(see Eq. 8.20 in Rohlfs \& Wilson \cite{rohlfs96})}, where
$\theta_{\rm a}^{''}$ and $\theta_{\rm b}^{''}$ indicate the Full
Width at Half Power (FWHP) linewidths of the major and minor axes of
the elliptical synthesized beam in arcseconds.

\section{Results}

\subsection{Continuum emission}
Fig.\,\ref{continuum} shows the $\lambda \sim$1.3\,cm continuum of
IC\,342, Maffei\,2, and NGC\,253. We estimate that the calibration
accuracy of the continuum is $\sim 10\%$.  The 23.7 GHz continuum
emission of IC\,342 peaks toward $\alpha_{J2000}=03^{\rm h}46^{\rm
m}47\fs85$, $\delta_{\rm J2000}=68^\circ05'46\farcs0$ with a total
flux density of 33\,mJy 
and a maximum of 7.1\,mJy/beam. The emission is resolved with an
apparent FWHP extent of $9\farcs 2 \times 5\farcs 6$ (major $\times$
minor axis), which corresponds to
a deconvolved source size of $8\farcs4 \times 5\farcs 0$ at a P.A. of
$+78^{\rm o}$. This extent is compatible with the observations by
Turner \& Ho (\cite{turner83}), who separated the thermal from the
non-thermal emission. The thermal 6\,cm continuum emission determined
by these authors is $\sim$40\,mJy for IC\,342. Considering this flux
density and a 
spectral index $\alpha$ of $-$0.1 (S$_\nu \propto \nu^\alpha$), the
expected thermal emission is $\sim$ 36\,mJy at 23.7\, GHz, which is
consistent with our measured value of 33 mJy.

The 23.7 GHz continuum emission from Maffei\,2 is centered at
$\alpha_{\rm J2000}=02^{\rm h}41^{\rm m}55\fs02$, $\delta_{\rm
J2000}=59^\circ36'17\farcs2$ with a total flux density of 40\,mJy
and a maximum of 8.2\,mJy/beam. The emission is resolved with an
apparent FWHP extent of $9\farcs 6 \times 5\farcs 0 $ (major $\times$
minor axis), which corresponds to
a deconvolved source size of $9\farcs 1 \times 3\farcs 4$ at a P.A. of $\sim
0^{\rm o}$. This extent is compatible with the higher resolution VLA
2\,cm data by Turner \& Ho (\cite{turner94}), who found that the
extended continuum emission is dominated by non-thermal radiation.
They assumed an $\alpha$ of $-$0.8 for the extended continuum
in Maffei\,2.  The non-thermal component is, however, spatially
related to compact thermal sources. Taking the above mentioned
spectral index, the 
expected flux density is 31.8 mJy at 23.7 GHz, and is smaller than
our value of 40 mJy. This suggests that a significant fraction
of the 23.7 GHz continuum is due to thermal free-free emission.

Toward NGC\,253 we have detected 23.7\, GHz continuum emission
centered at $\alpha_{\rm J2000}=00^{\rm h}47^{\rm m}33\fs16$,
$\delta_{\rm J2000}=-25^{\rm o}17'17\farcs5$ with a total flux
density of 468\,mJy and a maximum of 179\,mJy/beam. This flux
density is compatible, within the estimated errors, with the data
obtained by Geldzahler \& Witzel (\cite{geldzahler}) and Ott et al.
(\cite{ott05b}). The emission is resolved with an apparent FWHP
extent of $5\farcs 1 \times 2\farcs 8 $ (major $\times$ minor axis),
which corresponds to a deconvolved 
source size of $4\farcs 2 \times 1\farcs 7$ at a P.A. of $45^{\rm o}$.
The spatial extent of this emission is compatible with the higher
resolution images obtained by Ulvestad \& Antonucci (\cite{ulvestad97}).

\subsection{The NH$_3$ emission}

Our maps of the NH$_3$ emission toward IC\,342, Maffei\,2, and
NGC\,253 confirm that the single dish ammonia spectra previously
detected toward these galaxies (Mauersberger et al. \cite{mau03}) are
composed of 
the contributions of different giant molecular clouds (GMCs).  In the
analysis of the data, for each identified giant cloud, Gaussians were
fitted to the ammonia spectra. Table \ref{linefit} contains the fit
parameters for each GMC for IC\,342,
Maffei\,2, and NGC\,253, respectively. The average linewidths for
IC\,342 are $\sim$ 40 km s$^{-1}$ while for Maffei\,2 and NGC\,253
they are larger by a factor up to four (see \S\ref{NGC253}). This, and
the low velocity resolution of $\sim$20 km s$^{-1}$ for IC\,342 and
Maffei\,2 and $\sim$40 km 
s$^{-1}$ for NGC\,253 does not allow for a deconvolution of the line
profiles into velocity sub-structures inside the GMCs or to deconvolve
the satellite lines as it has been done for Galactic 
center clouds (McGary \& Ho \cite{mcgary02}).

\begin{table*}[tp]
\caption{Gaussian fit parameters, NH$_3$ column densities, and
rotational temperatures} \label{linefit}
\begin{tabular}{l c c c c c c c}
\hline \hline   & Position & & & &  \\
  & $\alpha$(J2000)~~ $\delta$(J2000)  &NH$_3$& $S$  &FWHM  & $v_{\rm hel}$ &$N(J,K)$&$T_{\rm rot}(1,1;2,2)$ \\
      & (h m s) ~~~~($^\circ$ ' '') &($J,K$) &(mJy) & (km~s$^{-1}$) & (km~s$^{-1}$)  &10$^{13}$cm$^{-2}$&K\\

\hline
   &   {\bf IC\,342}$^a$ & &   & \\
A  & 03 46 48.47 +68 05 43.2   &(1,1)&  1.9 & 39.0  & 23.6&13.3 &35.8\\
 &    &(2,2)& 1.3  & 40.5 & 21.5 &7.1\\
C  & 03 46 49.10~ +68 05 50.9   &(1,1)& 5.6  & 44.0 & 50.3&16.1&31.4 \\
  &     &(2,2)& 3.7  & 40.2 & 46.9 &7.3\\
D  & 03 46 49.60~ +68 05 58.6  &(1,1)& 3.1  & 45.1  & 52.2 &5.1&24.7\\
  &    &(2,2)& 1.4  & 42.2 & 50.4&1.6 \\
E  & 03 46 47.47 +68 05 41.8  &(1,1)& 2.2  & 37.4 & 13.7 &16.5&25.0\\
   &          &(2,2)& 0.9  & 39.4 & 13.2 &5.3\\
   \multicolumn{2}{l}{\bf Total Map Flux}&(1,1) &\multicolumn{2}{c}{1.43$^b$} &  & &29.3 \\
 & &(2,2)&\multicolumn{2}{c}{0.89$^b$} & &  &  \\

\hline
 &  {\bf Maffei\,2} & & &   & \\
A & 02 41 55.66 +59 36 26.0 &(1,1)& 2.0 & 65.6  & $-$82.6 &15.5&29.0\\
  & &(2,2)& 1.3 & 54.3  & $-$82.8 &6.25\\
B$^c$ & 02 41 55.20~ +59 36 20.0 &1,1)& 4.8 & 74.2 & $-$84.9&17.0&47.8\\
  & &(2,2)& 3.8 & 88.3 &  $-$85.9 &12.0\\
C & 02 41 54.89 +59 36 10.2 &1,1)& 5.6 & 66.1 & 0.15&15.6&23.9\\
 & &(2,2)& 4.0 & 37.1 & $-$1.9&4.7 \\
D & 02 41 53.94 +59 36 59.3 &1,1)&7.3 & 57.2 &  23.5&9.5&33.0 \\
 & &(2,2)& 5.4 & 49.6 & 18.1&4.56 \\
 \multicolumn{2}{l}{\bf Total Map Flux}&(1,1) & \multicolumn{2}{c}{1.83$^b$}& & & 32.8\\
 & &(2,2)& \multicolumn{2}{c}{0.78$^b$}& &  &  \\
\hline & {\bf NGC\,253}& & & & \\
A & 00 47 32.05 $-$25 17 26.5 &(3,3)& 22.2  & 85.0 & 313.7&74 \\
B & 00 47 32.33 $-$25 17 20.0 &(3,3)& -- & -- & -- &\\
C & 00 47 32.81 $-$25 17 21.1 &(3,3)& 26.5 & 62.2 & 290.7&122 \\
D & 00 47 33.25 $-$25 17 16.6 & (3,3)&17.8 & 161.6 & 210.9 &170\\
E & 00 47 33.66 $-$25 17 12.8 &(3,3) &14.7 & 79.1 & 187.4 &123\\
F & 00 47 34.06 $-$25 17 11.2 &(3,3) &14.8 & 69.6 & 214.9 &62\\
\multicolumn{2}{l}{\bf Total Map Flux}&(3,3) &\multicolumn{2}{c}{11.8$^b$} & & & \\
\hline
\hline
\multicolumn{8}{l}{$^a$ The nomenclature follows the one presented by
Downes \etal (\cite{downes92}, their Fig. 1).} \\
\multicolumn{8}{l}{~~~Component B is not seen in ammonia.} \\
\multicolumn{8}{l}{$^b$ $\int S {\rm d}v$ (Jy\, K\,km\,s$^{-1}$)} \\
\multicolumn{8}{l}{$^c$ There is $\sim$1\,mJy of absorption in the
NH$_3$ (2,2) line toward source B of Maffei 2 in the} \\
\multicolumn{8}{l}{~~~velocity range between $-20$ and 100\,km\,s${-1}$.
Since this absorption is at the edge} \\
\multicolumn{8}{l}{~~~of our band we did not fit a Gaussian to it.} \\
\end{tabular}\\
\end{table*}

\subsubsection{IC\,342}
\label{IC342}

Improvements in the performance of the VLA interferometer have
resulted in a highly increased dynamic range for our 
images of the ammonia emission from IC\,342 as compared to Ho et
al. (\cite{ho90}).  Our channel maps of the ammonia (1,1) and (2,2)
emission (Figure \ref{ic342-ch}) reveal clear substructure not
apparent in previous measurements. The ammonia emission is present
between 0.4 \kms ~and 80 \kms~.

\begin{figure*}
 \begin{tabular}{c c}
 \multirow{2}{11.5cm}{\epsfig{file=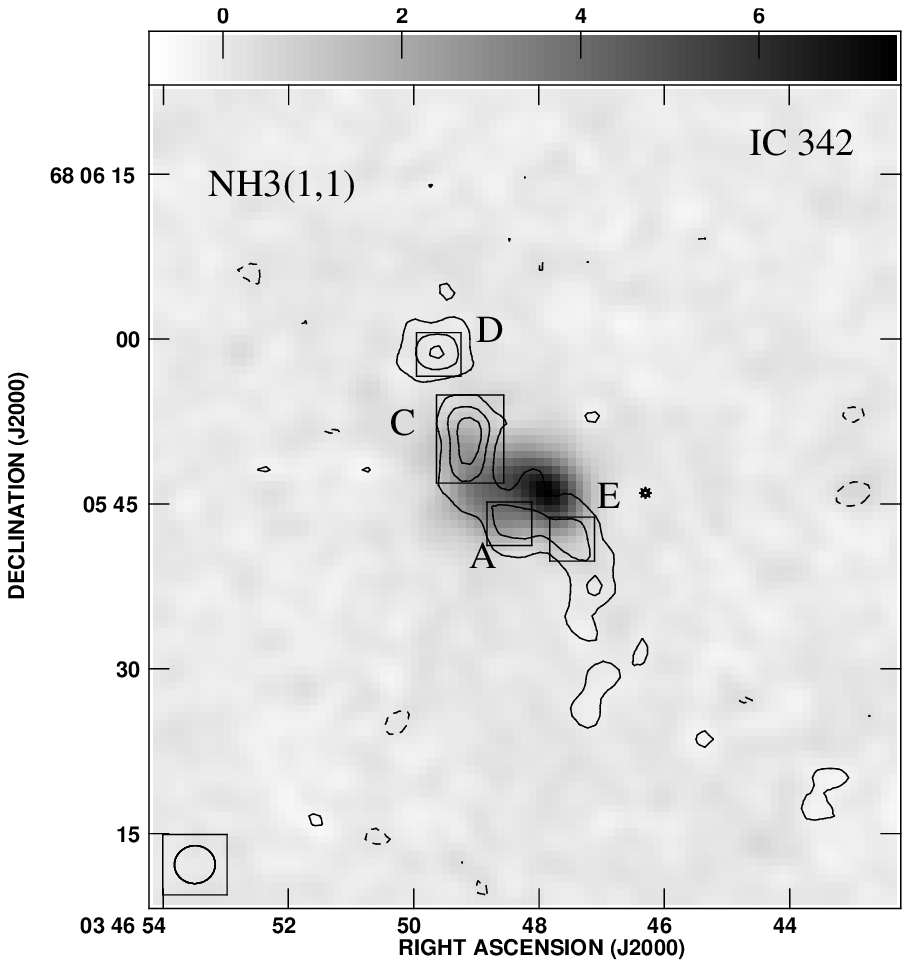,width=5.5cm}
  \epsfig{file=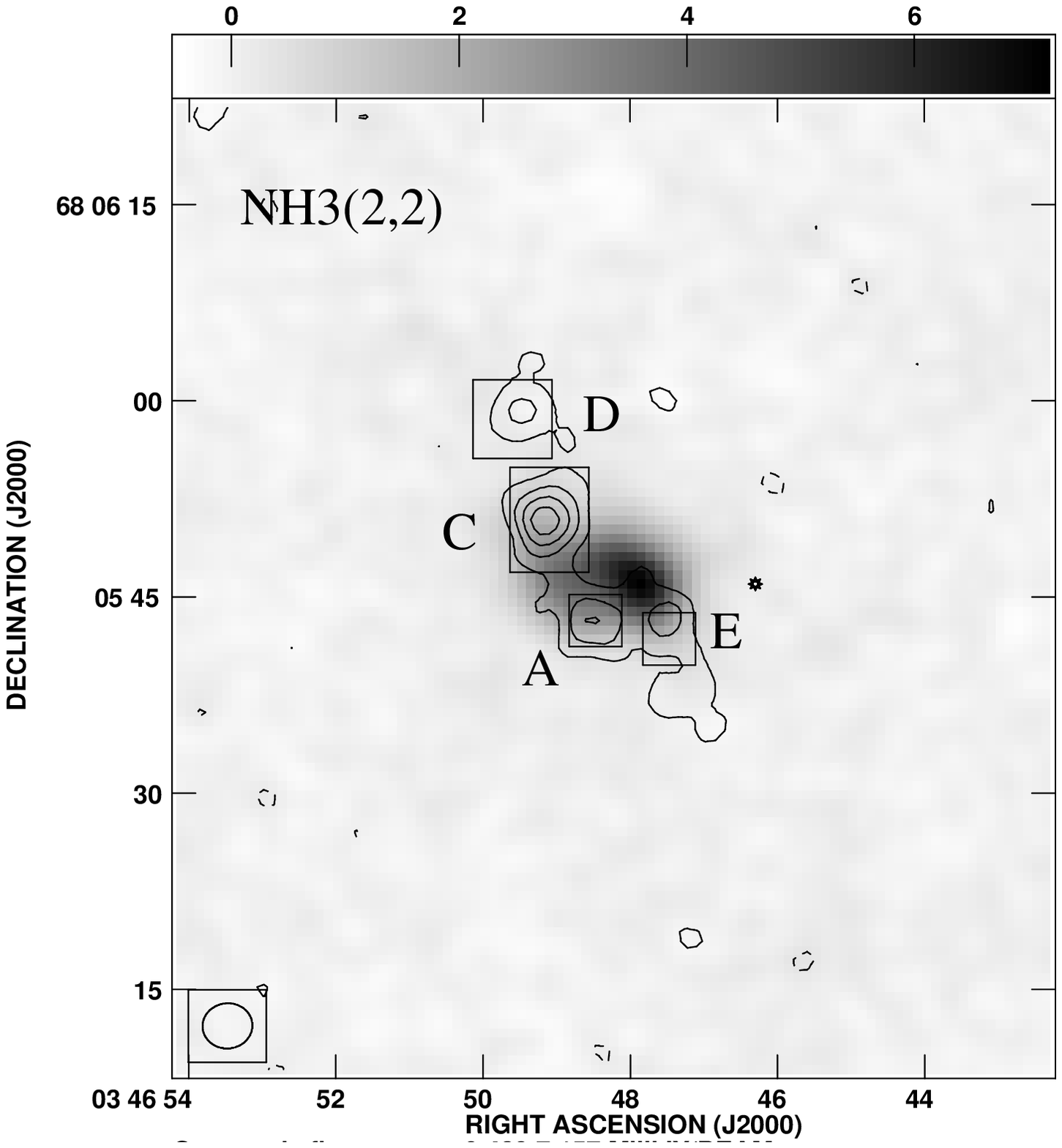,width=5.5cm}}&
\epsfig{file=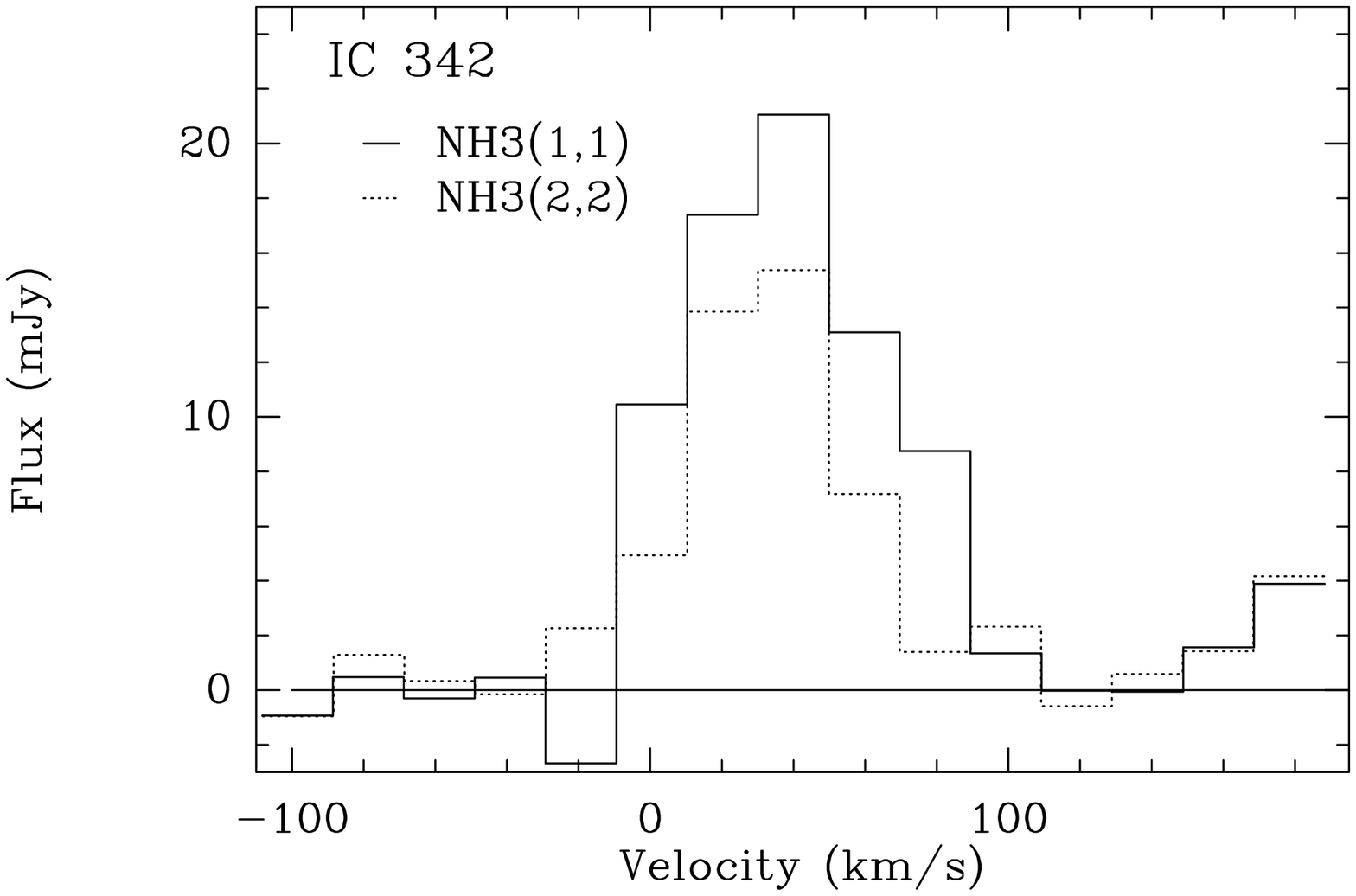,width=3.5cm,angle=-90}\\ &
  \epsfig{file=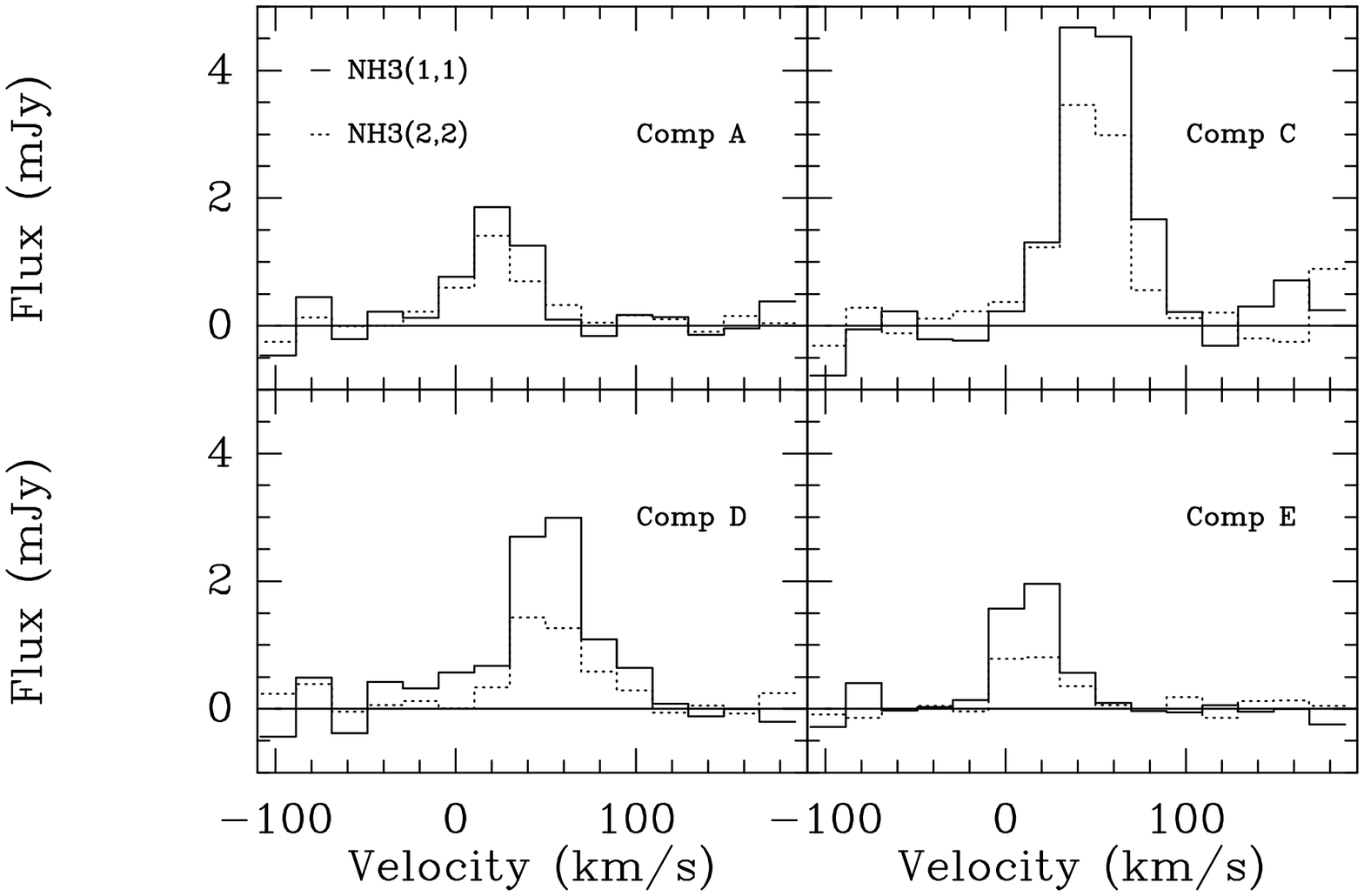,width=4.5cm,angle=-90}\\
 \end{tabular}
 \caption{Shown as contours, emission integrated over the velocity
 range 0.45 to 59.8 km~s$^{-1}$ for the NH$_3$(1,1) transition
 ({\it{left}}) and NH$_3$(2,2) transition ({\it{central}}) toward
 IC\,342. The contour levels are -3, 3, 6, 9 and 12 times 0.7 mJy
km~s$^{-1}$ beam$^{-1}$ for the (1,1) map and -3, 3, 6, 9 and 12
times 0.5 mJy km~s$^{-1}$ beam$^{-1}$ for the (2,2) map. Overlaid in
 gray scale for each panel are the radio continuum maps at the
 respective frequencies ($\sim$23.7\,GHz) with the upper bars
 providing the flux density scale in mJy~beam$^{-1}$. An asterisk
 indicates the location of the (western) water vapor maser detected by
 Tarchi et al. (\cite{tarchi02}), while 
 the boxes indicate the areas that were taken for integrating the
 emission from each ammonia cloud identified from these data. The
 spectra to the right show the integrated NH$_3$ emission (upper
 panel) and the spectra toward the four regions depicted in the
 contour maps (lower panel).} \label{ic342maps}
\end{figure*}

Figure \ref{ic342maps} shows the velocity integrated emission for
both transitions. The ammonia emission is concentrated in four main
giant clouds. The peak position of each cloud is given 
in Table \ref{linefit}, while the corresponding spectrum is shown in 
Figure \ref{ic342maps}. Each spectrum was fitted with a single
Gaussian in order to obtain the central velocity and the apparent
linewidth.

In the regions selected around some of the GMCs (see
Figure~\ref{maffei2map}), the overall
velocity integrated fluxes of the (1,1) and (2,2) 
transitions are 0.54 and 0.30\,Jy\,km\,s$^{-1}$. This is significantly
smaller than the total flux in these lines contained in our maps
(1.43\,Jy\,km\,s$^{-1}$ for the (1,1) line and 0.89\,Jy\,km\,s$^{-1}$
for the (2,2) line) and indicates that the regions selected represent
only about 35\% of the total NH$_3$ emission. Ammonia is therefore not
restricted to the main identifiable giant molecular clouds but is also
detected from the intercloud gas. Within the calibration uncertainties
and accounting for potential differences due to different beam sizes,
the measured total flux is similar to the single dish fluxes from
Mauersberger et al. (\cite{mau03}) which are 1.1\,Jy\,km\,s$^{-1}$ for
the (1,1) line 
and 1.24\,Jy\,km\,s$^{-1}$ for the (2,2) line. Our interferometric data are
therefore not missing significant amounts of single dish flux from the
ammonia emission.

\subsubsection{Maffei\,2}

\begin{figure*}
 \begin{tabular}{c  c}
  \multirow{2}{11.5cm}{\epsfig{file=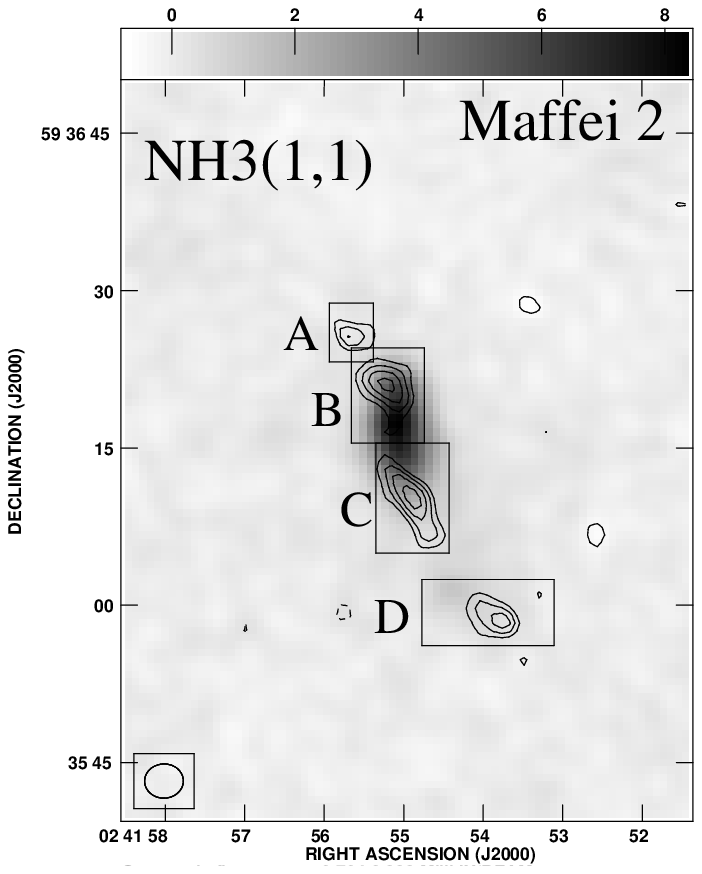,width=5.5cm}
  \epsfig{file=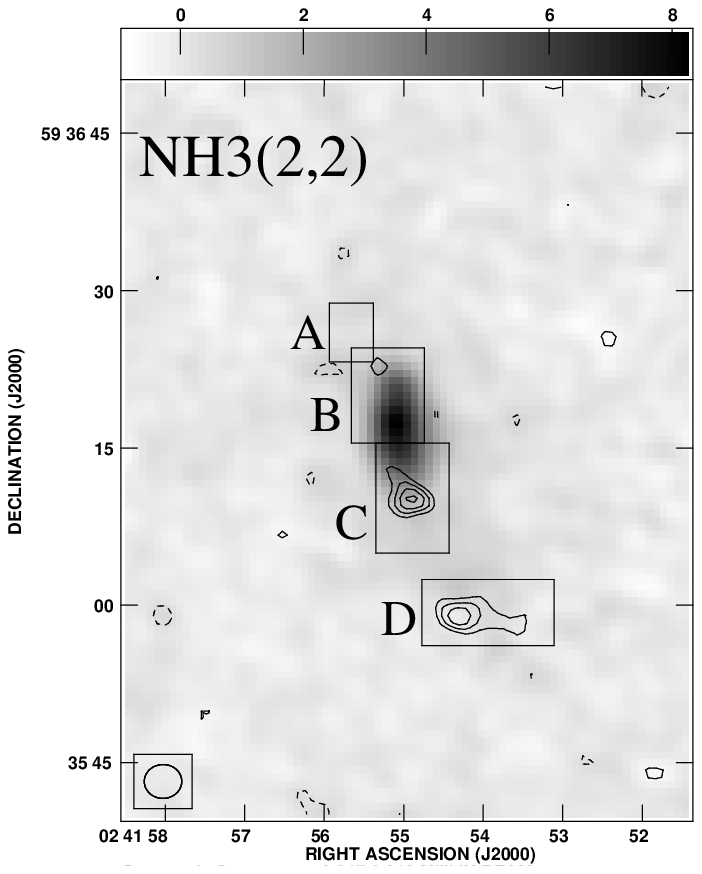,width=5.5cm}}&
  \epsfig{file=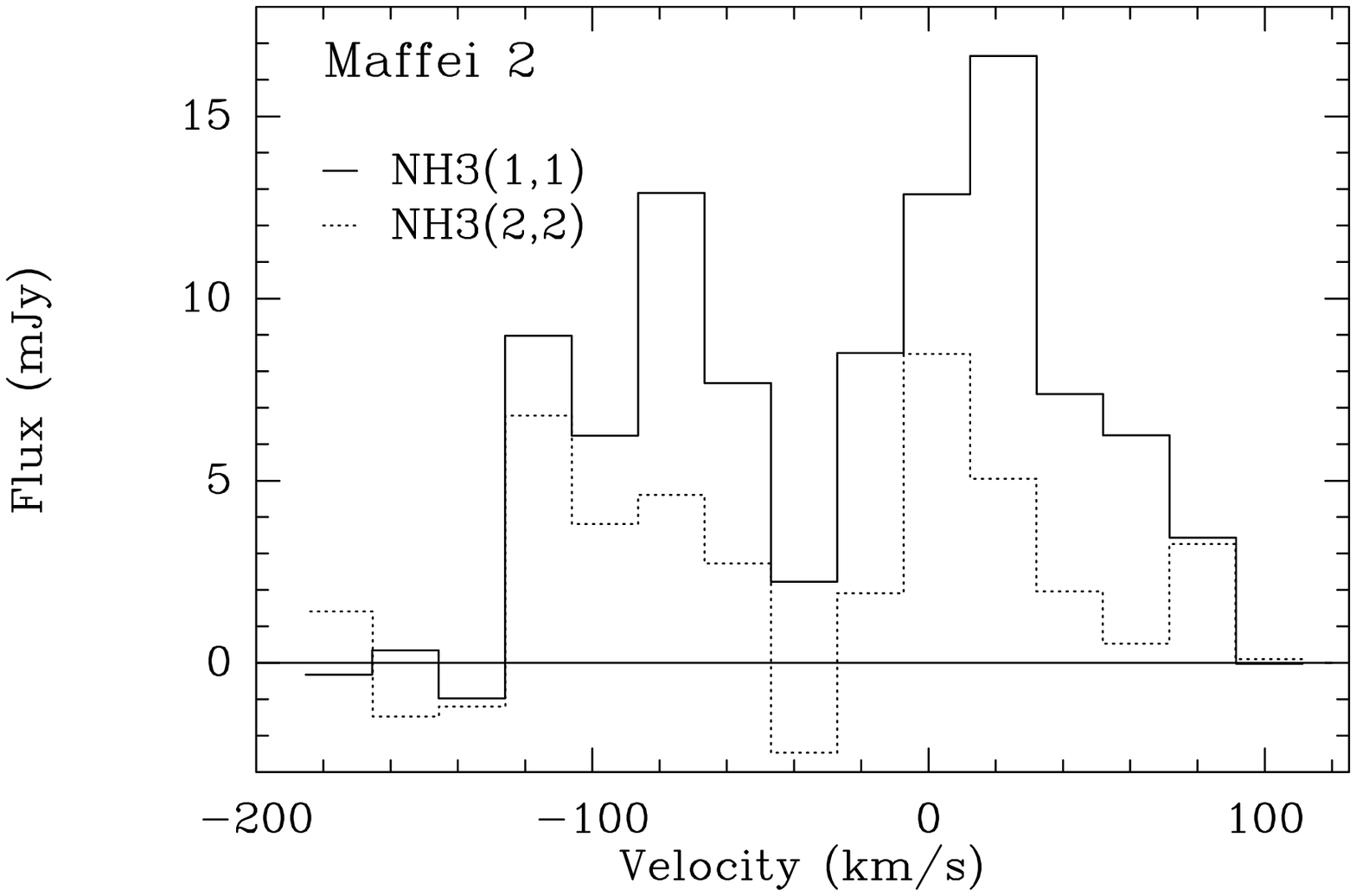,width=3.5cm,angle=-90}\\&
  \epsfig{file=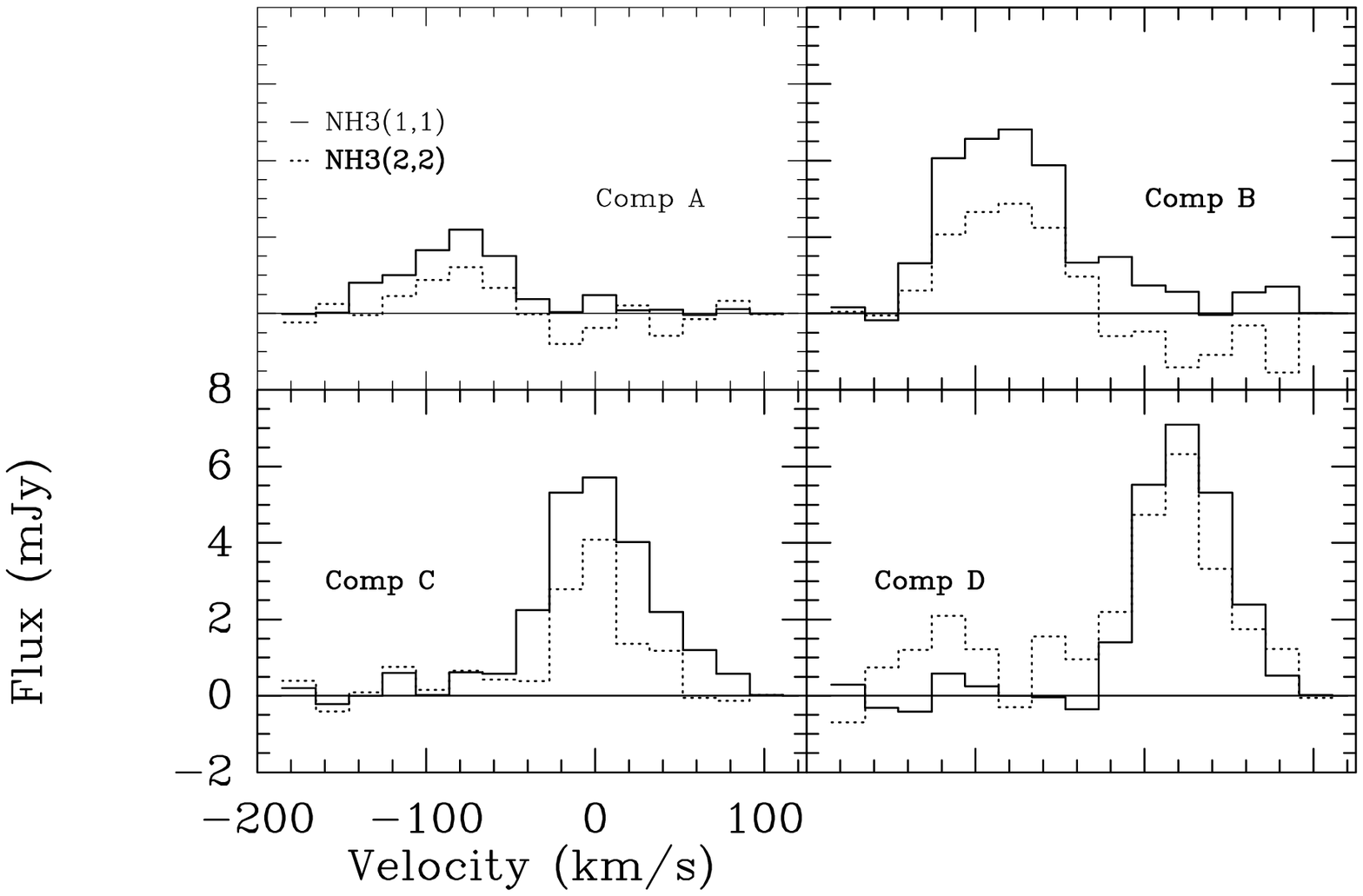,width=4.5cm,angle=-90}\\
 \end{tabular}
 \caption{Images of the velocity-integrated NH$_3$(1,1) and
  NH$_3$(2,2) lines from Maffei\,2.  The emission was
  integrated from $-$135 km~s$^{-1}$ to 62 km~s$^{-1}$.  The
  gray scale images are the continuum maps at the respective
  frequencies ($\sim$23.7\,GHz) with the upper bars providing the flux
  density scale in mJy~beam$^{-1}$. The contour levels are $-3$, 3, 4,
  5, and 6 times 1.3 
  mJy km~s$^{-1}$ beam$^{-1}$ for the (1,1) map and -3, 3, 4, 5, and 6
  times 1.8 mJy km~s$^{-1}$ beam$^{-1}$ for the (2,2) map. The spectra
  to the right show the integrated NH$_3$ emission (upper panel) and
  the spectra toward the four regions depicted in the contour maps
  (lower panel).} \label{maffei2map}
\end{figure*}

The ammonia (1,1) and (2,2) spectral channel measurements of Maffei\,2
are shown in Figure~\ref{maffei2-ch}, revealing velocity structure
which forms four main giant clouds along a ridge extending from the
north-east to the south-west (Figure~\ref{maffei2map}).  There
is emission toward all 4 peaks. In addition, in peak B, which is close
to the continuum emission peak, the NH$_3$ 
(2,2) is seen in absorption against the 1.3 cm continuum in the
velocity range between -20 and 100 km\,s$^{-1}$.  Maffei\,2 is located
close to the Galactic plane and the line of sight material causes a
visual extinction of 5.6 mag (Fingerhut et
al. \cite{Fingerhut2007}). The NH$_3$ absorption we observe is most
likely from gas 
within Maffei\,2 and not from intervening gas in our own Milky Way
since a) the observed lines are much wider than a few km\,s$^{-1}$ and
b) the velocities expected from gas in this quadrant of the Milky Way
would be at negative velocities.  In
Table \ref{linefit} we give the results of Gaussian fits toward the
peaks A-D. The fluxes of the (1,1) and (2,2) emission lines 
contained in the regions A---D are 1.3 and 0.8 Jy\,km\,s$^{-1}$, while
the total line flux from these transitions contained in our map is 1.8
and 0.8 Jy\,km\,s$^{-1}$. For the (1,1) line this indicates that the
emission from regions A---D represents 70\% of the total flux in the
mapped area. Such a percentage may also be representative for the
NH$_3$ (2,2) emission regions, but a direct comparison is not straight
forward due to the mixture of emission and absorption features in this
line, which are difficult to disentangle.
The single dish fluxes from
the 40$''$ Effelsberg beam are 1.2 and 1.1 Jy\,km\,s$^{-1}$ (Henkel et
al. \cite{hen00}). For the (1,1) line this is smaller than the total
flux in our map, which is expected since the NH$_3$ emitting region in
Maffei\,2 is not small compared to the Effelsberg beam. In particular,
peak D is right at the half power level of the Effelsberg beam.

\subsubsection{NGC\,253} 
\label{NGC253}

\begin{figure*}
\begin{tabular}{c  r}
\multirow{2}{10cm}{\includegraphics[bb=10 20 320
300,width=10cm,clip]{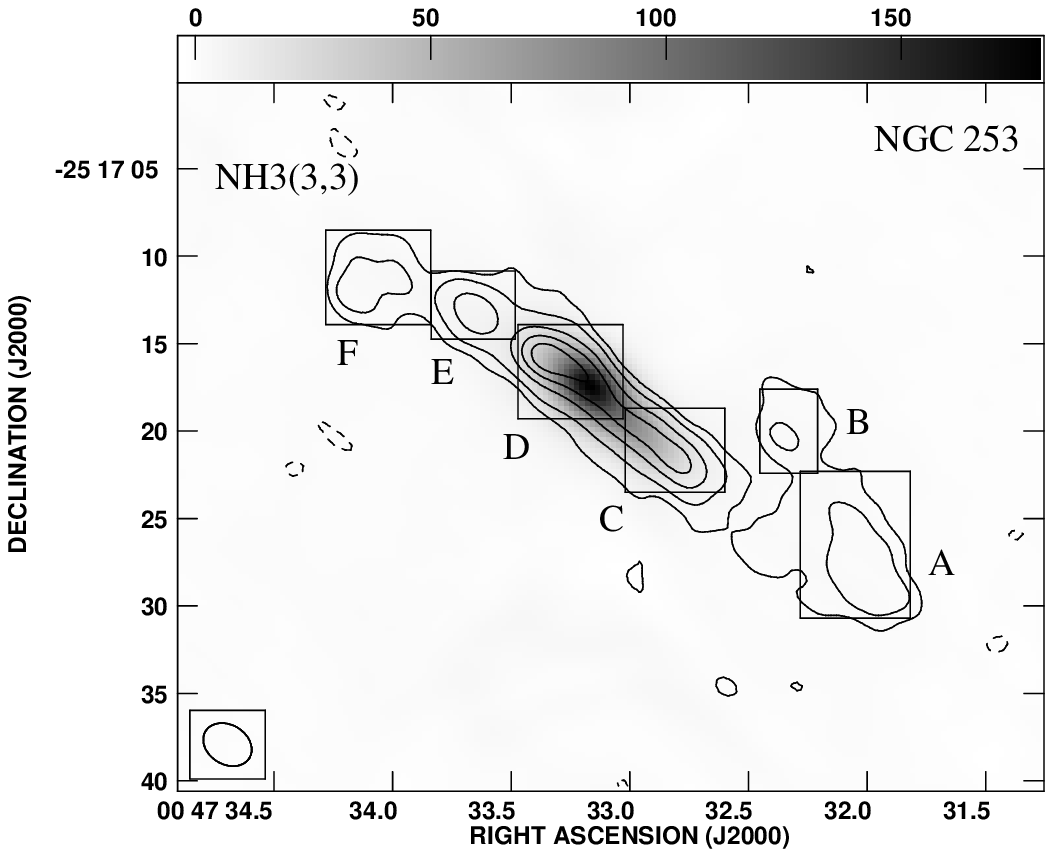}}&
\includegraphics[bb=30 120 500 600,width=6.7cm,angle=-90,clip]{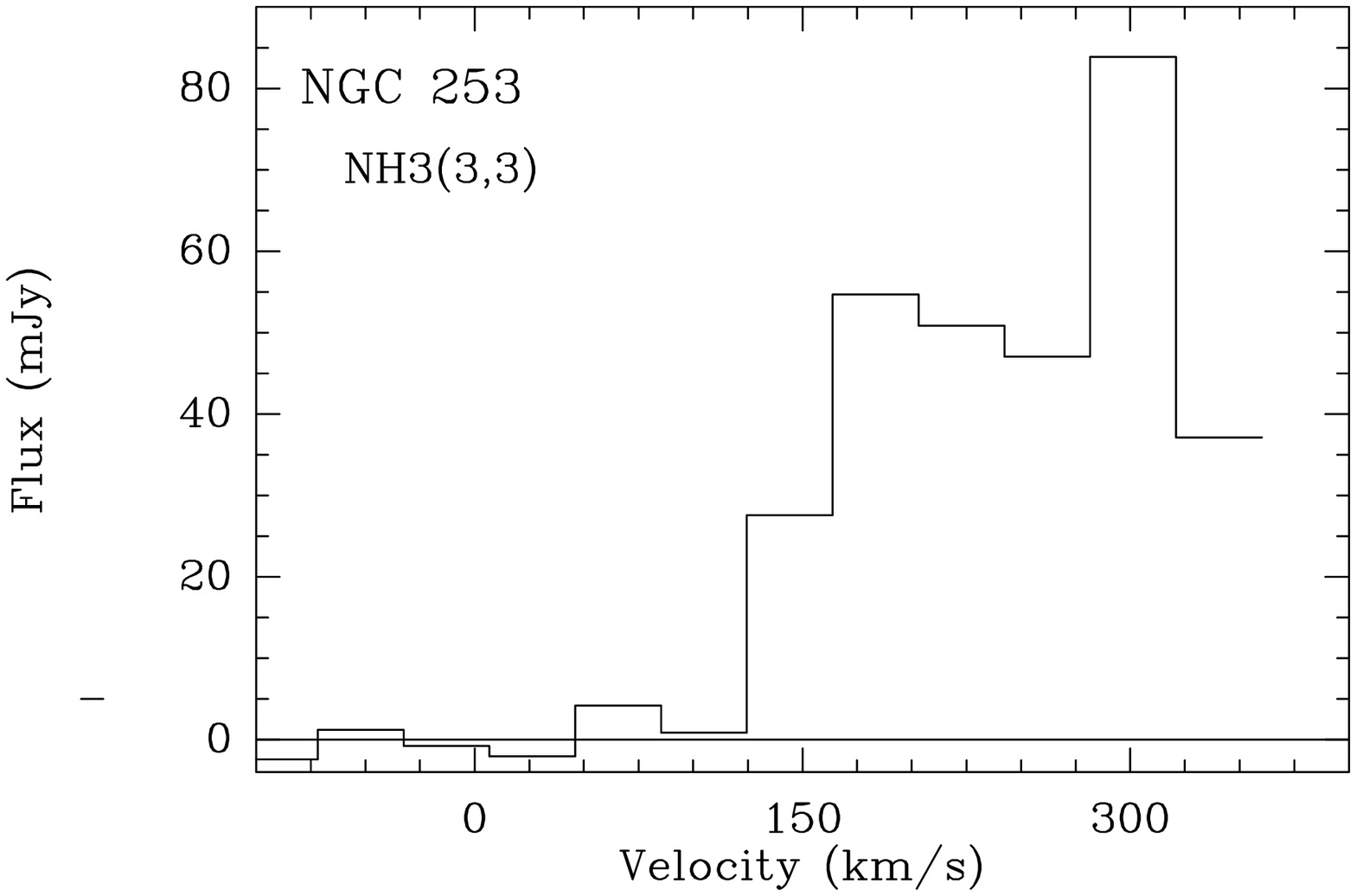}
\\ & \includegraphics[bb=124 96 470
600,width=5cm,angle=-90,clip]{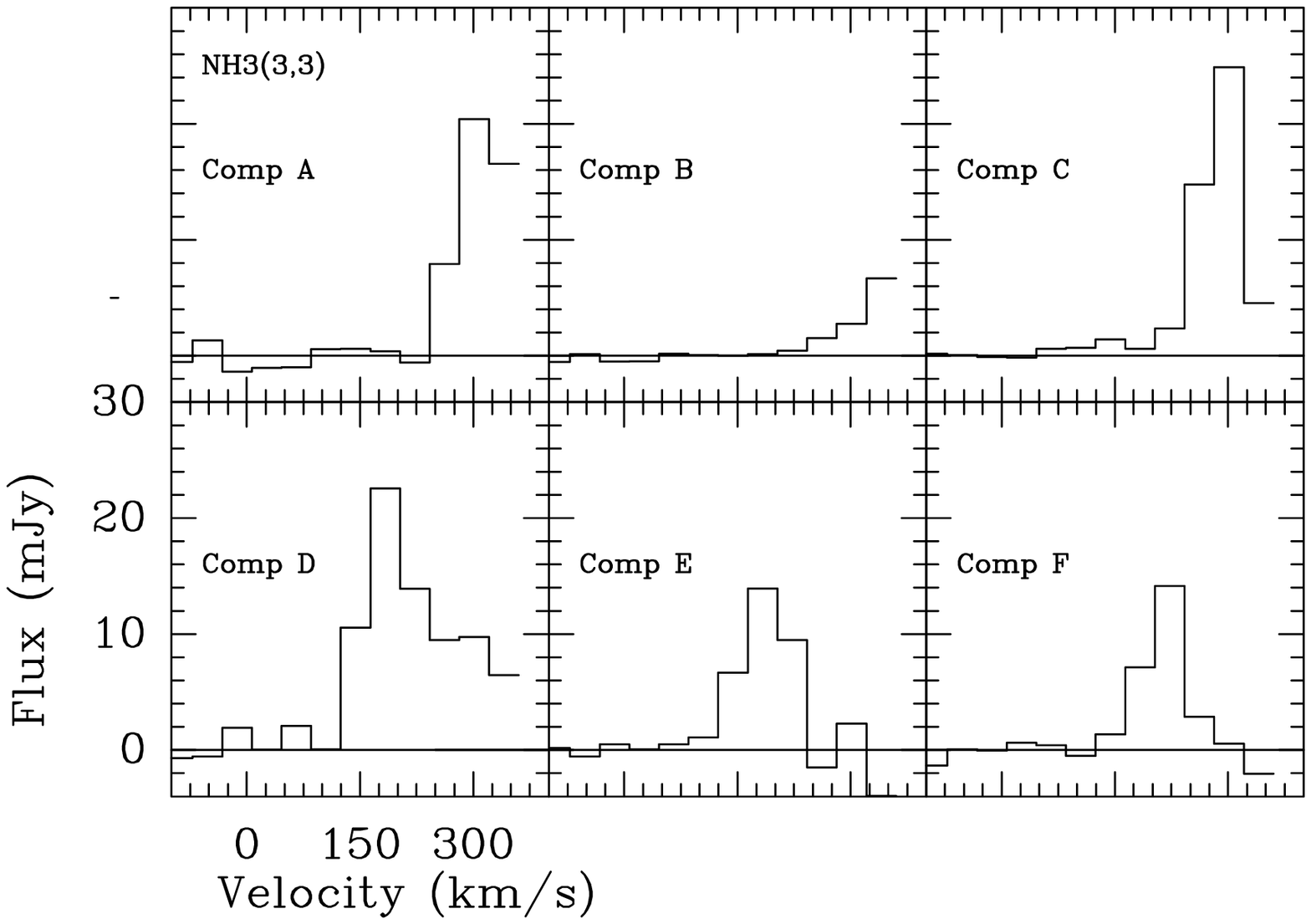} \\
\label{ngc253todosispec}
\end{tabular}\\
\caption{\label{ngc253map} {\it Left frame}: Contours show the
velocity-integrated NH$_3$(3,3) transition emission from NGC\,253
integrated from 144\,km\,s$^{-1}$ to 341\,km\,s$^{-1}$.  Overlain is
the gray-scale image of the continuum at 23.7\,GHz with the upper bar
providing the flux density scale in mJy\,beam$^{-1}$. The contour
levels are $-$3, 3, 6, 12, 
18, 24 times 1.2 mJy~km~s$^{-1}$~beam$^{-1}$. {\it Right upper
frame}: Integrated NH$_3$ (3,3) spectrum. {\it Right lower frame}:
Ammonia (3,3) transition spectra integrated over the six clouds of
NGC\,253 identified by the boxes in the velocity integrated NH$_3$
(3,3) emission (left panel).}
\end{figure*}

The left frame of Fig.\,\ref{ngc253map} shows the velocity-integrated
emission of the NH$_3$(3,3) transition. With an overall extent of
$40''\times 5''$ along a P.A. of $\sim 45^{\rm o}$, the NH$_3$
emission is much more extended than the 1.3\,cm continuum, which is
located near the ``center of gravity'' of the NH$_3$ emission. The
upper right of Fig.\,\ref{ngc253map} shows the integrated spectrum of
the NH$_3$(3,3) emission. The line shape and the flux of 8.6
Jy\,km\,s$^{-1}$ is only slightly smaller than 
the total flux detected in our maps (11.8\,Jy\,km\,s$^{-1}$), which is
similar to the single dish spectrum  
observed by Mauersberger et al (\cite{mau03}) with a 40$''$ beam,
which contains a flux of 10.9 Jy\,km\,s$^{-1}$. The flux measured by
Ott et al. (\cite{ott05b}) with the ATCA interferometer (synthesized
beam sizes $19'' \times 5''$ and $30'' \times 5''$) is
8.1 Jy\,km\,s$^{-1}$. Takano et al.
(\cite{Takano05a}) report a significantly higher value that is 30\%
higher than ours, namely 15.3\,Jy\,km\,s$^{-1}$.  This may be to a
very small part due to our 
incomplete spectral coverage of the NH$_3$ (3,3) line, but mainly
reflects the calibration uncertainties at such a low elevation.  We
will assume in 
the following that, as for Maffei 2, most of the flux density has been
recovered by our measurements.

The NH$_3$ emission arises from six giant clouds, that we label from A
to F with increasing right ascension. The lower-right 
frame of Fig.\,\ref{ngc253map} shows the NH$_3$(3,3) spectra
observed toward each of the six giant clouds (indicated by boxes in
the contour map). Each spectrum was fitted with a single Gaussian (see
Table \ref{linefit}). Typical linewidths of the
individual clouds are of order 70---160 km\,s$^{-1}$. This is
larger than the velocity separation of the hyperfine (HF) components
of the NH$_3$(3,3) transition. Under conditions of Local Thermodynamic
Equilibrium (LTE) and optically thin line emission, the outer
hyperfine components have an intensity 2.6\% of that of the main
component. Therefore, the fitted linewidths should well correspond to
the intrinsic linewidths of the clouds.  Channel maps of the ammonia
(3,3) emission are shown in Figure \ref{ngc253-33ch}. The ammonia
emission is present between 144\,km\,s$^{-1}$ and 341\,km\,s$^{-1}$.
The latter is the upper limit to the velocity extent of the emission
dictated by observational bandwidth limitations (see \S\ref{ObsDataRed}).

\section{NH$_3$ column density and rotational temperature}

The total column density of the two states of an ammonia $(J,K)$ inversion
doublet is given by 

\begin{equation}
N(J,K)=1.65\,10^{14}{\rm
cm^{-2}}\nu^{-1}\left[J(J+1)/K^2\right] \Delta v T_{\rm ex}\tau_{\rm tot},
\label{eq:njk}
\end{equation}

\noindent{where} $\nu$ is the inversion frequency in GHz, $\Delta v$
is the FWHP linewidth in km\,s$^{-1}$, $T_{\rm ex}$ is the excitation
temperature in K, and $\tau_{\rm tot}$ is the sum of the peak optical
depths of the five groups of hyperfine (hf) components comprising each
inversion transition.  For the $(J,K)$=(1,1) transition, $\tau _{\rm
tot}$=2.000$\tau_{\rm m}$ (m: main group of hf components); and for 
the (2,2) and (3,3) transitions $\tau_{\rm tot}$=1.256$\tau_{\rm m}$,
and 1.119$\tau_{\rm m}$, respectively.  For higher-lying inversion
transitions $\tau_{\rm tot} \simeq \tau_{\rm m}$. Assuming that the 
background continuum can be neglected, the brightness temperature of
the main group of hyperfine components $T_{\rm b}$ of an ammonia
transition relates to its excitation temperature as $T_{\rm b}=T_{\rm
ex}(1-\exp (-\tau_{\rm m}))$, i.e. in the optically thin limit $T_{\rm
b}\sim T_{\rm ex}\tau_{\rm m}$. Using the transition parameter that
can be directly observed, namely the main-beam brightness temperature
$T_{\rm mb}$, instead of $T_{\rm b}$ results in beam averaged
(instead of source averaged) column densities.

The transition optical depths are unknown and cannot be easily determined
since Doppler broadening masks individual hyperfine
satellite emission.  Takano et al. (\cite{takano2005b}) claim low
optical depths from line-to-continuum ratios of absorption lines
toward Arp\,220. The latter is, however, also affected by the unknown
source covering factor of the molecular gas. From observations of
ammonia emission toward molecular clouds toward the Galactic Center
region, where linewidths are of order 10\,km\,s$^{-1}$,
H\"uttemeister et al. (\cite{huettemeister1993}) derive optical
depths of about two for the main group of hyperfine satellites
of the (1,1) and (2,2) transitions. They also find higher optical depths
toward the stronger peaks of molecular emission. The Galactic Center
data presented by H\"uttemeister et al. (\cite{huettemeister1993})
were obtained with a linear resolution of 1.7 to 3.2 pc, while the VLA
beam size toward the galaxies observed corresponds to about 50\,pc
(see \S\ref{Intro}). It
is plausible that the source-averaged opacities of the observed NH$_3$
transitions toward our galaxy sample are smaller than the value of
$\sim 2$ determined toward Galactic Center clouds, which would suggest
that $T_{\rm ex}\tau_{\rm tot}\sim T_{\rm mb}$.

The synthesized beam averaged $T_{\rm mb}$ values are associated with
the spectral flux density, $S_\nu$, by the relation in
Equation~\ref{eq:jyperk}.
%
%
With these assumptions, beam averaged column densities can be
derived using

\begin{equation}
N(J,K)=\kappa10^{16}{\rm cm^{-2}}(\theta_{\rm a}'' \theta_{\rm b}'')^{-1
}S_{\rm Jy}\Delta v ({\rm km\,s^{-1}}),
\label{columndensityeq}
\end{equation}

\noindent{where} $\kappa=6.04$, 2.81, 2.21, and 1.47 for the
$(J,K)$=$(1,1)$, $(2,2)$ and $(3,3)$ and $(6,6)$ transitions discussed
in this paper, respectively.

The rotational temperature between any $(J,J)$ and ($J',J'$) levels
can then be determined using 

\begin{equation}
T_{\rm rot} = {\Delta E \, ({\rm
K})/ {\ln\left[{(2J+1)g_{\rm op}\over (2J'+1)g'_{\rm
op}}{{N(J',J')}\over{N(J,J)}}\right]}}.
\label{eq:trot}
\end{equation}

\noindent{Here} $g_{\rm op}$ is 2 for  ortho ammonia  ($K=3,6$), and 1
for para ammonia ($K=1,2$).


\section{Discussion}

\subsection{IC\,342}

\subsubsection{Distribution and excitation of the molecular gas}

Distance estimates toward the nearly face-on and highly obscured
spiral galaxy IC\,342 range from 1.8 Mpc (McCall \cite{mccall89})
to 8 Mpc (Sandage \& Tammann \cite{sandage74}).  As a consequence of
this distance 
ambiguity the size and luminosity of this galaxy have been classified
as similar to or much in excess of the Galactic values. Recently 
Tikhonov \& Galazutdinova (2010) derived a distance of
3.9$\pm$0.1 Mpc using stellar photometry. To facilitate a comparison
with previous studies, here we adopt the distance proposed by Saha et
al. (\cite{saha2002}), 3.3\,Mpc.  The giant molecular clouds in the
central bar of IC\,342 (e.g. Sakamoto et al. \cite{sakamoto99}) have a
linear size which is similar to that of the Sgr~A and Sgr~B2
molecular clouds in the center of our Galaxy. The inner 400~pc of
IC\,342 have the same far infrared luminosity as the inner 400~pc of
our Galaxy and the 2$\mu$m luminosities indicate that the central
regions contain similar masses of stars (e.g. Downes et al. \cite{downes92}).
Fortunately, IC\,342 is viewed face-on, and its nuclear region
is therefore much less obscured than that of the Milky Way, although IC\,342 is
located behind the Galactic plane ($b^{\rm II}$ $\sim$
10$^{\circ}$).


\subsubsection{Comparison of NH$_3$ with other tracers of dense gas in IC\,342}

According to Downes et al. (\cite{downes92}) and Sakamoto et al.
(\cite{sakamoto99}), the highest concentrations of CO and HCN are
present in a region of diameter $\sim$40$''$ near the nucleus of
IC\,342. Downes et al. (\cite{downes92}) discussed the presence of
at least five molecular complexes, the most interesting of which is
``Cloud B'' (following the notation of Downes et al.). Only this
region is associated with powerful H\,{\sc ii} regions, while the
other complexes appear to be too hot and too turbulent to form stars
with similar efficiency. The nucleus and Cloud B are only
$\sim$3$''$ apart.  The ammonia emission is in the form of four
unresolved or poorly resolved maxima along an elongated curved
structure of 40$''$ (350\,pc) length. Four of these maxima can be
identified with the HCN maxima A, C, D, and E observed by Downes et al.
(\cite{downes92}). The ammonia ridge continues southward of HCN peak E
without resolving single GMCs. No NH$_3$ emission can be seen
toward ``Cloud B''. 

When we compare our NH$_3$ (1,1) and (2,2) emission with the various
interferometric molecular emission maps by Meier \& Turner
(\cite{meier01,meier05}) and Usero et al. (\cite{usero06}), we find the best
correspondence (including the extension south of cloud E) with the
distributions of C$^{18}$O $(1-0)$, 
CH$_3$OH (methanol) ($2_k-1_k$) and N$_2$H$^+$ emission.  We also find
excellent correspondence with images of H$^{13}$CO$^+$ $(1-0)$ and SiO
$(2-1)$, although in these 
Plateau de Bure maps the extensions south of Cloud E are not visible
as they are outside the primary beam.  There is also correspondence
between HCN peaks A, C, D, and E and HNCO ($4_{04}-3_{03}$),
although the extension south of 
Cloud E is not seen in HNCO. With HNC ($1-0$), as with HCN ($1-0$) (Downes
et al. \cite{downes92}), there is good coincidence of peaks A, C, D,
and E; however, Cloud B is seen in HNC and HCN but (as already
mentioned) not in NH$_3$. The 
poorest coincidence with our NH$_3$ images is with the
distributions of CCH $1-0;3/2-1/2$ and C$^{34}$S $2-1$, both of which
peak between Cloud A and the continuum peak.

Of the more complex molecules that coincide the best, namely
NH$_3$, CH$_3$OH and HNCO, all have in common that they are thought to
be liberated from grains by shocks. In order to be released from
grain mantles without being destroyed,  slow ($v_{\rm shock}=10-15
\rm km\,s^{-1}$) shocks are required (see the discussion in Mart\'in
et al. \cite{martin06a}).  Ammonia, methanol and HNCO are easily
destroyed by dissociating radiation (Hartquist et
al. \cite{hartquist95}, Le Teuff et al. \cite{leteuff2000}). SiO, on
the other hand would need fast shocks ($v_{\rm shock}\ga 15-29 \rm
km\,s^{-1}$) to be released from grain cores (see the discussion in
Usero et al. \cite{usero06}).  Alternatively it has been proposed that
SiO can be generated in a high temperature environment in the gas
phase (Ziurys et al. \cite{ziurys89}). The good coincidence of SiO
with the slow shock tracers methanol, ammonia and HNCO would support
this latter interpretation. For all the mentioned tracer molecules,
our peaks A, C, D and E can clearly be identified, and they all seem to
avoid the continuum peak indicating the absence (or effective
shielding) of dissociating radiation.

In contrast, C$^{34}$S, CCH (and also to a certain degree HCN and
HNC) are strongly enhanced toward peak B. In the case of CS, this
can be explained by a chemical enhancement of the CS abundance of up
to $>10^{-6}$ due to the presence of a small amount of ionized sulfur
(Sternberg \& Dalgarno \cite{sternberg95}). Also the CCH abundance
is predicted to be enhanced in photodissociation regions (see the
discussion in Meier \& Turner \cite{meier05}), which would explain its
existence toward the only peak associated with H\,{\sc ii} regions.

\subsubsection{Kinetic temperature in the IC\,342 clouds}

In Table\,\ref{linefit} we list the derived rotational temperatures
between the (1,1) and (2,2) rotational levels for the four peaks
detected in these transitions; values range between 25\,K (Clouds D
and E) and 36\,K (Cloud A).  Since the rotational temperatures between
the (1,1) and (2,2) levels represent only a lower limit to the kinetic
temperatures it is interesting to compare these with those derived using
higher excitation transitions.

\begin{figure}
\includegraphics[width=8cm]{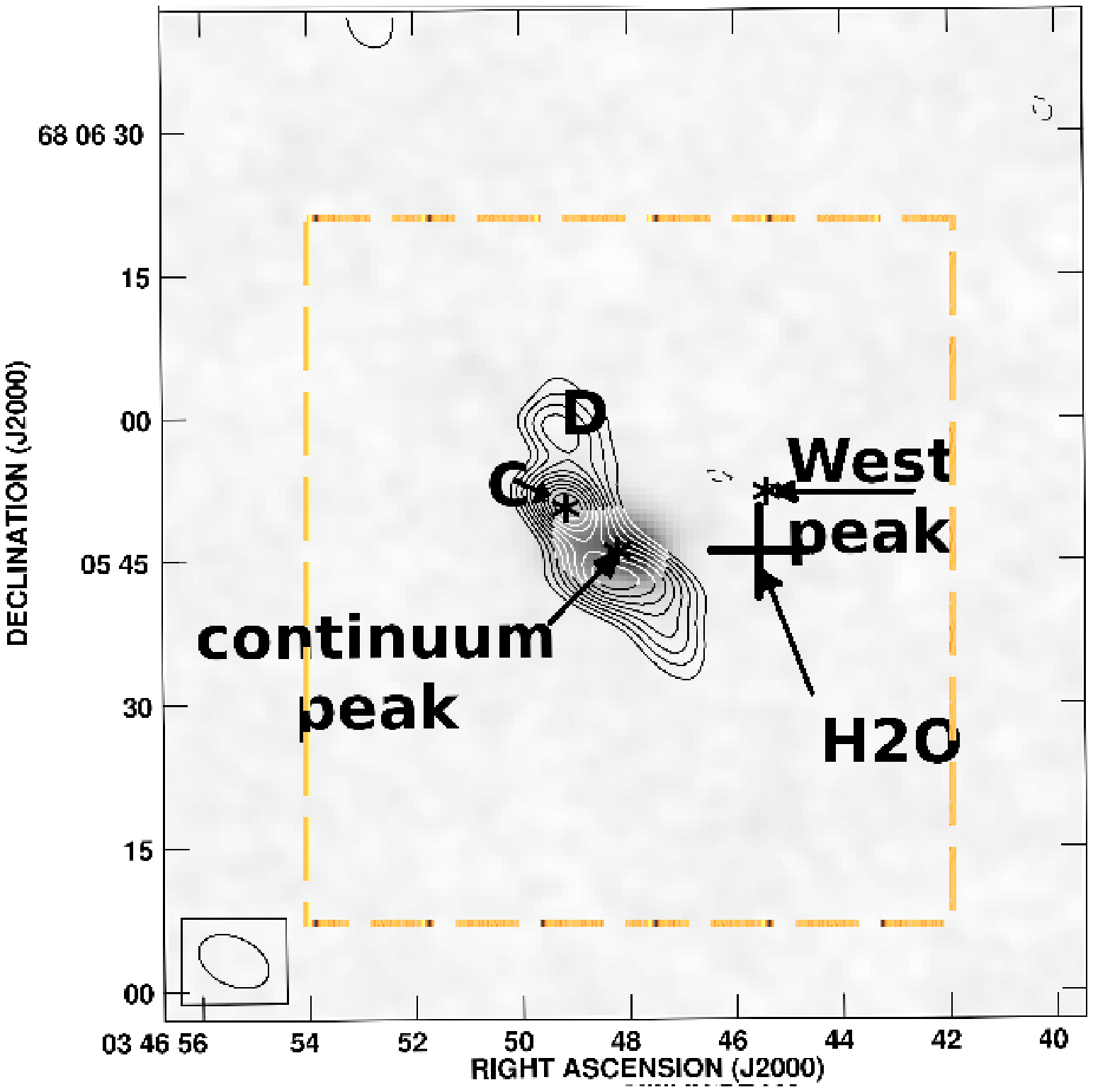}\\
\includegraphics[width=5cm, angle=-90]{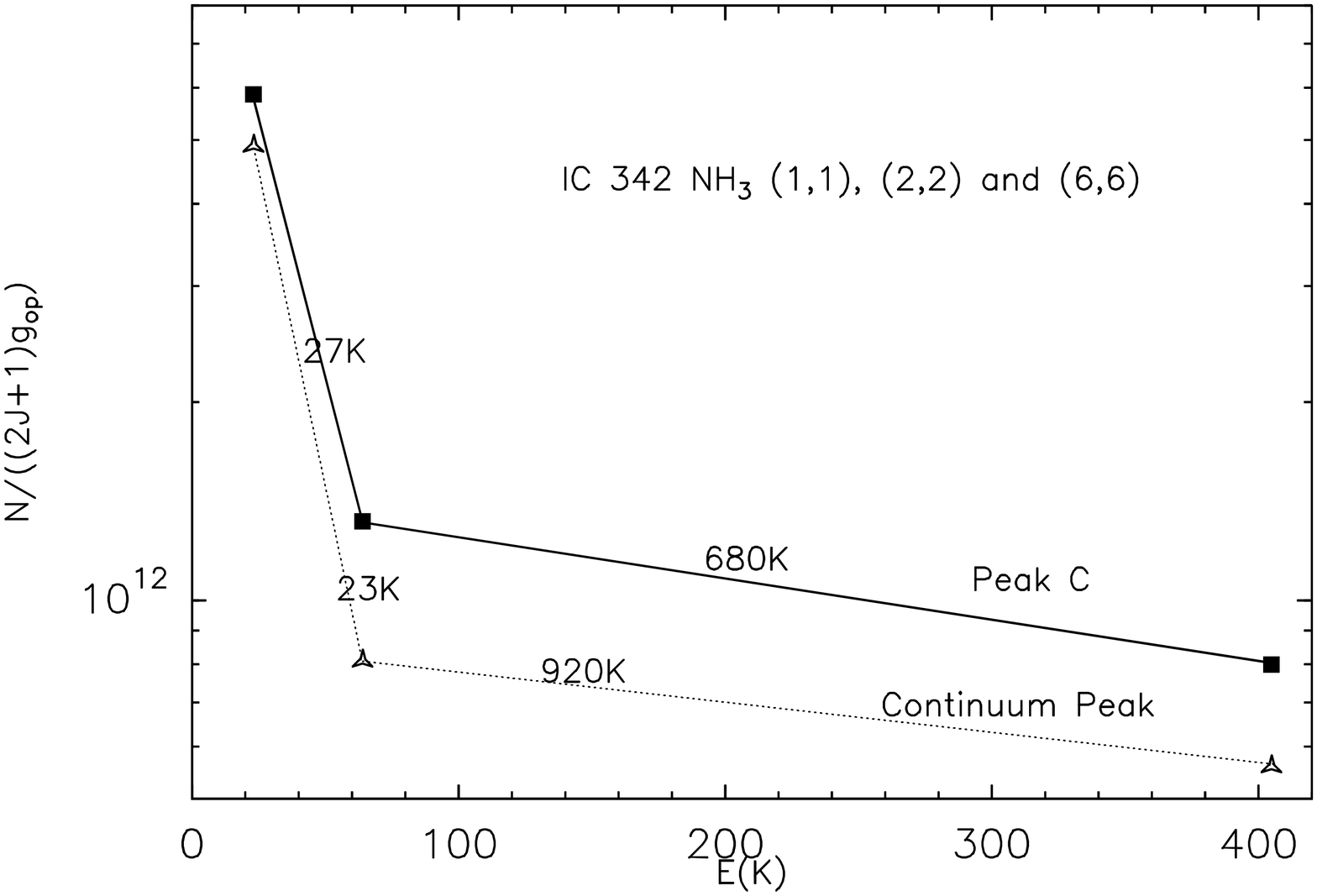}
\caption{{\it Upper frame}: Contours showing our IC\,342 NH$_3$
$(J,K)=(2,2)$ image smoothed to the same resolution as the
Montero-Casta\~no et al. (\cite{montero2006}) NH$_3$(6,6) emission
image. The gray scale shows the 1.3\,mm continuum, and the asterisks
denote the three (6,6) ammonia peaks. The H$_2$O maser detected by
Tarchi et al. (\cite{tarchi02}) is indicated as a cross, while the box
indicates the area depicted in Figure~\ref{ic342maps}. {\it Lower
frame}: Boltzmann plot for IC\,342 peak C and the "continuum" peak
using the measured (1,1), (2,2) and (6,6) transitions of NH$_3$. 
\label{ic342multiammonia}}
\end{figure}

The VLA map of the $(J,K)=(6,6)$ (E$_u$ = 411 K)
transition toward IC\,342 by Montero-Casta\~no et
al. (\cite{montero2006}) allows for a direct comparison with our images
of the lower excitation NH$_3$, and thus to study the excitation
conditions in the various giant molecular clouds of IC\,342.
Comparing the maximum of the spectral flux density, $S$, in their
(6,6) VLA map with Effelsberg data by Mauersberger et al.
(\cite{mau03}), Montero-Casta\~no et al. (\cite{montero2006})
conclude that no flux is missing in their NH$_3$ (6,6) image, i.e. the
three peaks represent indeed 100\% of the (6,6) emission. The
situation is different if one compares the flux density integrated
over the line profiles (i.e. $\int S$\,d$v$). The total
flux in the VLA NH$_3$ (6,6) image is 242 mJy\,km\,s$^{-1}$, while the 
Effelsberg spectrum presented by Mauersberger et al. (\cite{mau03})
contains a flux of 330 mJy\,km\,s$^{-1}$, i.e. 27\% of the NH$_3$
(6,6) line flux is missing in the VLA (6,6) map, most of which is
probably in the line wings. 

In Fig.\,\ref{ic342multiammonia} we show a version of our NH$_3$
(2,2) map smoothed to the same angular resolution as the (6,6)
ammonia map of Montero-Casta\~no (\cite{montero2006}). In the NH$_3$
(6,6) map, which has a resolution of $7\farcs 8\times 5\farcs 0$, three peaks
can be identified: The main NH$_3$ (6,6) peak can be identified with our
peak C within an accuracy of 1''. The second NH$_3$ (6,6) peak appears
to be north of our peaks A and E by 3'', south of the HCN peak B, and
coincides with the 
1.3\,cm continuum. It is remarkable that this second peak does not
coincide well with most other molecular line distributions presented
in Meier \& Turner (\cite{meier05}), except the HNC ($1-0$) and the
HC$_3$N $(10-9)$ transitions. The third peak, the ``Western Peak''
observed in NH$_3$ is the most mysterious since it does not coincide
with any published molecular distribution, including our NH$_3$ (1,1)
and (2,2) images, nor with any prominent continuum features. A notable
exception is a 1.3\,cm water maser discovered by Tarchi et
al. (\cite{tarchi02}; see \S\ref{WesternPeak}), whose positional
uncertainty is $5''$. The NH$_3$ (2,2) peak
D, on the other hand, has no counterpart in NH$_3$ (6,6).

In Fig.\,\ref{ic342multiammonia}b we show Boltzmann plots toward the
other NH$_3$ (6,6) peaks. For this comparison the (1,1) and (2,2)
images were smoothed to the same resolution as the (6,6) image. This
has the consequence that there is also some weak (1,1) and (2,2)
emission from the continuum peak.  We can
observe the same behavior as was noted in the various transitions
measured with a single dish toward IC\,342 by Mauersberger et al
(\cite{mau03}).  While the rotational temperatures between the (1,1)
and (2,2) transitions are between 23\,K (continuum peak) and 27\,K (peak
C), the (6,6)/(2,2) rotational temperatures are 680\,K toward peak C
and 920\,K toward the continuum peak. These values are higher than the
single dish rotational temperature of 440\,K fitted by Mauersberger
(\cite{mau03}) to the NH$_3$ (5,5), (6,6) and (9,9) data. In a
warm environment, radiative transfer simulations predict higher
$T_{\rm rot}$ values for the higher-excitation metastable energy levels,
gradually approaching $T_{\rm K}$, even if the gas is characterized by
a single kinetic temperature (e.g. Walmsley \&
Ungerechts \cite{walmsley83}, Danby et al. \cite{danby88}, Flower et
al. \cite{flower95}).  Alternatively, there may be several gas
components with different temperatures within our beam.

It was noted in \S\ref{IC342} that within the selected
regions our NH$_3$ (1,1) and (2,2) transitions 
recover less than 50\% and 75\% of the (1,1) and (2,2) single dish
fluxes, respectively.  This will not significantly alter the resulting
rotational and/or kinetic temperatures derived from these
measurements. The different molecular complexes we have discovered are
not much larger than the synthesized beam width of the VLA. There may,
though, be an extended emission component within which the measured
components are embedded. Furthermore, since the percentage of missing
flux is larger 
for the (2,2) than for the (1,1) transition, the rotational
temperatures given in Table~\ref{linefit} are lower limits.

\subsubsection{The nature of the ``Peak C'' and the ``Continuum
peak''}

The NH$_3$ emitting gas in the central region of IC\,342 should
originate from highly shielded regions since NH$_3$ has a low
dissociation energy and can be easily destroyed by a strong energetic
radiation field (Saha et al. \cite{Suto1983}). The temperatures
observed are comparable to the
kinetic temperatures of $\ga 600\,\rm K$ found in absorption toward
the H\,{\sc ii} regions in Sgr\,B2 (H\"uttemeister et al.
\cite{hue95}; Wilson et al. \cite{wilson2006}). The hot gas in Sgr\,B2
is of comparatively low 
density ($< 10^3\,\rm cm^{-3}$). H\"uttemeister et al.
(\cite{hue95}) conclude that the gas toward Sgr\,B2 is probably
heated by C-type shocks arguing that this kind of shock is able to
liberate large quantities of NH$_3$ from dust mantles and heat
them to high temperatures without destroying them. Like in the case
of Sgr\,B2 the CH$_3$OH and SiO observed in IC\,342 might be
associated with the hot gas component, at least for peak C and the
continuum peak.

\subsubsection{The nature of the ``Western Peak''}
\label{WesternPeak}

Toward the western peak, no definite (6,6)/(2,2) temperature can be
derived due to our non detections of the lower excitation transitions of
NH$_3$. A similar case where there is a strong presence of NH$_3$
(6,6) emission with a corresponding lack of emission in the lower
excitation transitions of NH$_3$ has also been observed toward our
Galactic center (Herrnstein \& Ho \cite{herrnstein2002}), although on
a much smaller linear scale. This lack of NH$_3$ (1,1) or (2,2)
emission toward an NH$_3$ (6,6) emission peak has been explained by a
geometric configuration whereby the lower-excitation transitions
experience foreground absorption by cool material along the line of
sight.  However, although this scenario might explain the absence of lower
excitation NH$_3$ transitions, it would be difficult to explain why one
does not see {\it any} other molecular emission toward this
position. Alternative mechanisms that can explain the
presence of (6,6) emission but not lower lying transitions include:
\begin{itemize}
\item An overabundance of ortho ammonia (i.e. with $K=0,3,6, \ldots$) by
a factor of $>2$ due to its formation in a very cold  (i.e. $<< 23\,K
$) environment.  Unfortunately, this configuration has never been
observed in any source in the Galaxy. The hypothesis could be
tested by observing the NH$_3$(3,3) transitions toward the
NH$_3$ (6,6) peak. 
\item A more viable hypothesis is that we are observing maser emission
in the (6,6) transition but not in the (1,1) and (2,2)
transitions. Such a maser as observed toward the Galactic star
formation region NGC\,6334\,I (Beuther et al. \cite{beuther2007})
could be detectable even if the column  
density is comparatively low as the detection of $^{15}$NH$_3$
$(J,K)$=(3,3) maser emission toward a Galactic hot core suggests (Mauersberger
et al. \cite{mau86}). Ammonia maser activity has also been
proposed toward NGC\,253: Ott et al. (\cite{ott05b}) reported NH$_3$ (3,3)
in emission toward a position where other, higher and lower lying
transitions were seen in absorption. In order to prove or discard the
hypothesis of NH$_3$ (6,6) maser activity toward the western peak,
one has to perform higher resolution mapping and/or to study other
high excitation transitions of ammonia.
\end{itemize}

It is interesting that the western peak coincides, within the accuracy
of the measurements, with narrow
($\Delta v \sim 0.5\, \rm km\,s^{-1}$) water maser emission detected
by Tarchi et al. (\cite{tarchi02}). This water maser was found to be
time variable on scales of weeks. Its integrated luminosity of L$_{\rm
H_2O}=5\,10^{-3}$\,L$_\odot$ is comparable to that of the Sgr\,B2 star
forming region near the center of the Milky Way. While the H$_2$O
flux from Sgr\,B2 is spread over many velocity components, only a
narrow flaring feature was observed toward IC\,342. This is likely a
consequence of its much larger distance, only revealing an occasional
flare while sensitivity limitations do not allow us to detect the more
``quiescent'' components.  Tarchi et
al. (\cite{tarchi02}) found that toward the location of the IC\,342 water
maser (and, hence, toward the ``western peak'') there is a chain of
optically identified sources that appear to be H\,{\sc ii} regions.

\subsection{Maffei\,2}

\subsubsection{Distribution and excitation of the molecular gas}

The barred, highly inclined ($i=67^{\rm o}$) spiral galaxy Maffei\,2
at a distance of $\sim$ 3.3 Mpc (see the discussion in Meier et al.
\cite{meier08}) is one of the nearest starburst galaxies, and hence,
can be studied with high spatial resolution. A moderately strong
nuclear starburst has been observed in the infrared (Rickard \&
Harvey \cite{rickard83}) and in radio continuum emission (e. g.\
Turner \& Ho \cite{turner94}).  Its nuclear star
formation rate is similar to those of M~83 and NGC\,253 (Ho et al.
\cite{ho90a}), but its low Galactic latitude ($b^{\rm II} \sim -20^{'}$)
makes optical studies difficult (e. g.\ Buta \& McCall \cite{buta83}).

Maffei\,2 is a galaxy with strong molecular emission from its
central region. CO emission reveals a $\sim 220$\,pc long bar (Meier
et al. \cite{meier08}), which is distinct from the bar seen at large
scale in this galaxy. From the location of dense gas and star
formation near the intersections of $x_1-x_2$ orbits, Meier et al.
(\cite{meier08}) concluded that the starburst in Maffei\,2 is
triggered by dynamics.

\subsubsection{Comparison of NH$_3$ emission with other molecules}

Four molecular cloud complexes (designated in our
Fig.\,\ref{maffei2map} as NH$_3$\,A--D) can be seen in NH$_3$
(1,1), two of which are situated north of the nucleus and the other
two toward the south.  The NH$_3$ (2,2) emission is mainly seen toward
the south of the nucleus. The large scale distribution of the NH$_3$ 
(1,1)  emission in Maffei\,2 follows that of the isotopic CO and the
HCN ($J=1-0$) emission (Meier et al. \cite{meier08}).

If one goes into a detailed comparison of the various measured
molecular distributions, one can see important differences between
NH$_3$ and the other molecules. In $^{12}$CO $(1-0)$, and also in 
$^{13}$CO $(1-0)$, the emission is not only concentrated toward
molecular cloud complexes but also a more widespread diffuse component
can be seen. This might arise due to a higher opacity and a better
signal to noise ratio in these transitions. The rarer isotopic
substitution C$^{18}$O in its $J=1-0$ and $2-1$ transitions indicates, on
the other hand, a similar contrast as NH$_3$. 
The $2-1$
transitions are reliable tracers of the H$_2$ column density for a
large range of H$_2$ densities (Wilson \&
Mauersberger \cite{wilson90}).  In $^{13}$CO $2-1$, we can 
well distinguish the four NH$_3$ clouds. Unlike NH$_3$, CO also
shows strong emission toward the continuum peak. The detailed distribution of
C$^{18}$O $2-1$ differs from the distribution of the NH$_3$
and the $^{13}$CO emission.  This could be explained in terms of
large variations of isotopic abundances of $^{18}$O or $^{13}$C, contamination
by the continuum, or just insufficient signal-to-noise (D. Meyer,
priv. comm.). To conclude, the similar distributions of large scale
NH$_3$ emission with other tracers of H$_2$ suggests that we are
indeed observing a 
representative fraction of the dense gas in NH$_3$ and that the
observed NH$_3$ distribution is not just due to excitation effects 
or due to varying chemical abundances.

\subsubsection{Kinetic temperatures}

In Table\,\ref{linefit} we show rotational temperatures between the
(1,1) and (2,2) levels for the four peaks detected in these
transitions.  Derived rotational temperatures range from 24\,K
(Clouds C) to 48\,K (Cloud B).  Since the $T_{\rm rot}$
between the (1,1) and (2,2) levels represents only a lower limit to
$T_{\rm K}$, it is interesting to compare these measurements
to rotational temperatures derived from higher-excitation NH$_3$
transitions.  Unfortunately such higher transitions are only available
from single dish measurements (e.g. Henkel et al. \cite{hen00},
Mauersberger et al. \cite{mau03}, Takano et al. \cite{Takano05a}).
Radial velocity information, can, 
however, be used for a comparison.  From the (1,1) to (4,4)
transitions, Henkel et al. (\cite{hen00}) suggested that the
$-$80\,\kms component (Clouds A and B) should be
warmer than the component at $+6$\kms\ (Clouds C and
D). This is also what is indicated from our high resolution 
data. On the other hand the inclusion of (6,6) data by Mauersberger et
al. (\cite{mau03}) shows no difference in excitation for the very hot
gas. This indicates that although there is very warm gas (130\,K) at
$-80$ {\it and } at $+6$\,\kms, the gas toward the velocity component
at $+6$\,\kms, i.e. clouds C and D has, on average, lower
than average excitation, which could be due to an additional cooler component.

\subsection{NGC\,253}
\subsubsection{The distribution of NH$_3$ and other molecules}

The Sculptor galaxy NGC\,253 is a highly inclined ($i$ $\sim$
78$^{\circ}$) nearby ($D$ $\sim$ 3.5 Mpc, Recola et
al. \cite{Rekola2005}) barred Sc spiral. It is 
one of the brightest sources of far infrared emission beyond the
Magellanic Clouds and has been studied in detail at many
wavelengths. NGC\,253 hosts a nuclear starburst.  Radio continuum
measurements indicate that NGC\,253 contains a large number
of potential centrally located supernova remnants and H\,{\sc ii}
regions (Ulvestad \& Antonucci \cite{ulvestad97}; Brunthaler et
al. \cite{brunthaler2009}).

A nuclear bar has been imaged in the CO, HCN and HCO$^+$
emission lines (Sakamoto et al. \cite{sakamoto06}, Knudsen et al.
\cite{knudsen07}), and many other molecules have been detected in
NGC\,253 (see e.g. the 2\,mm line survey of Mart\'in et al.
\cite{martin06b}). Among these, five turned out to be particularly
important to reveal the peculiar physical conditions of the
starburst region in this galaxy: SO, SiO, HNCO, CH$_{3}$OH, and
CH$_{3}$CN (Henkel et al. \cite{henkel87}; \cite{mau91a};
Nguyen-Q-Rieu et al. \cite{nguyen91}; H\"uttemeister et
al. \cite{hue97}). The abundances of these five molecules have been
found to be particularly high in NGC\,253. This indicates that in
contrast to a ``late starburst'' such as M\,82, where the chemistry is
affected by photodissociation, the chemistry in NGC\,253 is dominated
by slow shocks which are able to release molecules from grains
without destroying them (see also Garc{\'{\i}}a-Burillo et
al. \cite{garciaburillo01}, Usero et al.
\cite{usero04}, \cite{usero06}). Ammonia belongs to the same
class of molecules, thought to be enriched in C-shock
environments and which is easily destroyed by dissociating radiation
due to its low dissociation energy (Suto \& Lee \cite{Suto1983}).

Takano et al. (\cite{Takano05a}) presented VLA maps of the (1,1),
(2,2), and (3,3) transitions of NH$_3$.  The data presented in this
paper only include the ammonia (3,3) transition but with a higher
spatial resolution ($2\farcs9 \times 2\farcs2$ instead of $4\farcs5
\times 2\farcs5$).  The agreement between the two (3,3) maps is
good.  Comparing this map with a super-resolved NH$_3$ map
(resolution: $5''\times 5''$) by Ott et al. (\cite{ott05b}) we find
excellent agreement with the locations of clouds A, B and C, while
clouds D, E and F are shifted between 2 to 5$''$ with respect their peak
positions.  This discrepancy can probably be explained by the
deconvolution process used by Ott et al. (\cite{ott05b}). There is
also good agreement between the features seen in our map and the 
$3\farcs 8\times 2\farcs 6$ resolution maps of HCN and HCO$^+$
presented by Knudsen et al. (\cite{knudsen07})

The individual GMCs in NGC\,253 are more pronounced in NH$_3$ than in
HCN or HCO$^+$. This may be due to the fact that the latter
transitions being optically thick, and therefore the
contrast between the 
clouds and the intercloud gas does not reflect the contrast in column
density of HCN or HCO$^+$. Another notable difference is that our
Cloud B is about 3$''$ north of the nearest HCN and HCO$^+$
feature. This HCN and HCO$^+$ feature, where no NH$_3$ is measured,
corresponds in location and velocity to one of the two 
``molecular superbubbles'' identified by Sakamoto et
al. (\cite{sakamoto06}), and is within the positional uncertainties
identical to the superbubble already seen in the NH$_3$ data of Ott et
al. (\cite{ott05b}). Also toward the other superbubble seen in the CO
data presented by Sakamoto et al. (\cite{sakamoto06}) and being
located in the northeastern edge of the measured NH$_3$
distribution (Fig.\,\ref{ngc253map}), weak HCN and
HCO$^+$ emission can be seen at the correct velocity, but no
NH$_3$. At the opposite edge of the NH$_3$ distribution, our Cloud A
is clearly visible in HCN, but not in HCO$^+$. 

The agreement of the NH$_3$ distribution with that of $^{12}$CO $(2-1)$,
$^{13}$CO $(2-1)$, $^{13}$CO $(3-2)$ and C$^{18}$O $(2-1)$ measured by
Sakamoto et al (\cite{Sakamoto2011}) 
is excellent. Taking into account our sightly coarser resolution we
can reproduce all the details in the distribution of those
transitions. Since CO isotopes are thought to be among the most
reliable tracers of the H$_2$ column density, this supports our notion
that NH$_3$ traces the bulk of the gas in the central molecular zone of
NGC\,253. The reason may 
be that at the high temperatures in this region the abundance of NH$_3$,
which otherwise can vary by several orders of magnitude due to freeze
out onto grains, is quite constant, since all NH$_3$ is in the gas
phase. The same may be true for the central molecular zone of the
Milky Way, which has 
similarly high temperatures to a degree that even SiO can be observed
throughout the region (Riquelme et al. \cite{Riquelme2010}).

To conclude, our NH$_3$ distribution follows closely that of
C$^{18}$O (2-1), which is thought to be an excellent indicator of 
molecular hydrogen. This suggests that the temperatures derived from
NH$_3$ are representative of the bulk of the molecular gas in the
central region of NGC\,253. HCN and HCO$^+$ also follow the general H$_2$
distribution. However, their emission is, with respect to NH$_3$,
enhanced toward regions where fast shocks and/or photodissociating
radiation can be expected.

\subsubsection{Ammonia maser emission or unusual ortho/para ratios?}

Takano et al. (\cite{takano02})  noted that the (3,3) transition was
stronger than expected toward NGC\,253, and concluded that an
anomalous ammonia ortho/para ratio was the cause. Also Ott et
al. (\cite{ott05b}) explained the unexpectedly high intensity of the
(3,3) transition toward the NE peak in terms of an ortho/para ratio for
ammonia of 2.5--3.5. On the other hand, Ott et al. (\cite{ott05b})
also mention that toward the
continuum position of NGC\,253 they measure the (3,3) transition
in emission, whereas all other transitions are observed in
absorption. This would suggest that maser emission in the (3,3)
transition is the cause of the higher than anticipated intensities,
which is amplifying the background continuum.

In our measurements of the integrated (3,3) spectrum (see
Fig.\,\ref{ngc253map}) the high velocity SW component (V
$\sim$300 \kms) is brighter than the lower velocity NE component
(V $\sim$160 \kms). This is contrary to the lower resolution data
presented by Ott et al (\cite{ott05b}) and Mauersberger et al (\cite{mau03})
where the 170 km\,s$^{-1}$ component is equal in intensity or stronger
than the 300\,km\,s$^{-1}$ component. Note that consistently in all
other ammonia transitions observed by Mauersberger et
al. (\cite{mau03}) and Ott et al. (\cite{ott05b}) the 
300\,km\,s$^{-1}$ component is the strongest. The flux density of the
170\,km\,s$^{-1}$ component is comparable in our measurements and
those presented by Ott et al.  The 300\,km\,s$^{-1}$ component, on the
other hand, is 50\% stronger in our measurements. From the channel
maps in Fig.\,\ref{ngc253map} and also 
from Takano et al. (\cite{Takano05a}), it appears as if this extra flux
could come from cloud C (clouds SW1 in the notation of Takano et
al. \cite{Takano05a}) which is close to the continuum peak. 

How can the different profiles be explained other than by calibration
uncertainties and errors? One possibility is time variable emission
on a time scale of a few years. If the (3,3) transition is masing this
would imply that the emission comes from a region small in angular
extent near the center of NGC\,253.  Such maser emission would also
imply that the intensity is very high within a very narrow transition
linewidth. 

Inconsistent sensitivity to spatial structure could be an alternative
explanation.  The situation here is that the higher resolution VLA map
results in a higher line flux than that derived from the Australia
Telescope Compact Array (ATCA) image presented by Ott et
al. (\cite{ott05b}). If there is some large scale absorption component
on top of the emission from the GMCs, this might be ``missed'' by the
VLA but not by the ATCA and may explain why the 300\,km\,s$^{-1}$
component is stronger with the VLA than with ATCA. Apparently higher
spatial and spectral resolution 
observations not only of the (3,3) transition but also of other ortho
or para transitions are necessary to settle this issue.

\section{Conclusions}

Imaging measurements of the NH$_3$ $(J,K)=(1,1)$ and (2,2) transitions
toward IC\,342 and Maffei\,2 and of the NH$_3$ (3,3) transition toward
NGC\,253 have been presented.  Derived lower-limits to
the kinetic
temperatures determined for the giant molecular clouds in the centers
of these galaxies are between 25 and 50\,K.  In general, there is good 
agreement between the distributions of NH$_3$ and other H$_2$ tracers,
suggesting that NH$_3$ is representative of the distribution of dense
gas in these galaxies.

For IC\,342, a comparison to high resolution NH$_3$ data indicates a
molecular component with kinetic temperatures as high as 700 ---
900\,K. Furthermore, the ``Western Peak'' of IC\,342, also known to host
an H$_2$O maser, is observed in the NH$_3$ (6,6) transition but not in
lower-excitation NH$_3$ transitions.  This is suggestive of maser
emission in the (6,6) transition. 

Comparing the (3,3) line profile from NGC\,253 with those obtained
from earlier lower resolution studies, the V $\sim$300\,km\,s$^{-1}$
component appears to be enhanced.  Explanations could involve a time
variable (3,3) maser or a large scale absorption component not seen by
us.


\begin{acknowledgements}
We thank the referee, Jean Turner, for excellent suggestions which
improved this work.  Rainer Mauersberger has been supported by a grant
(AYA2005-07516) of the Ministerio de Educaci\'on y  Ciencias of
Spain.\\
\end{acknowledgements}

\appendix

\section{Electronic tables and figures}
\begin{figure*}[p]
\begin{tabular}{c c c}
\epsfig{file=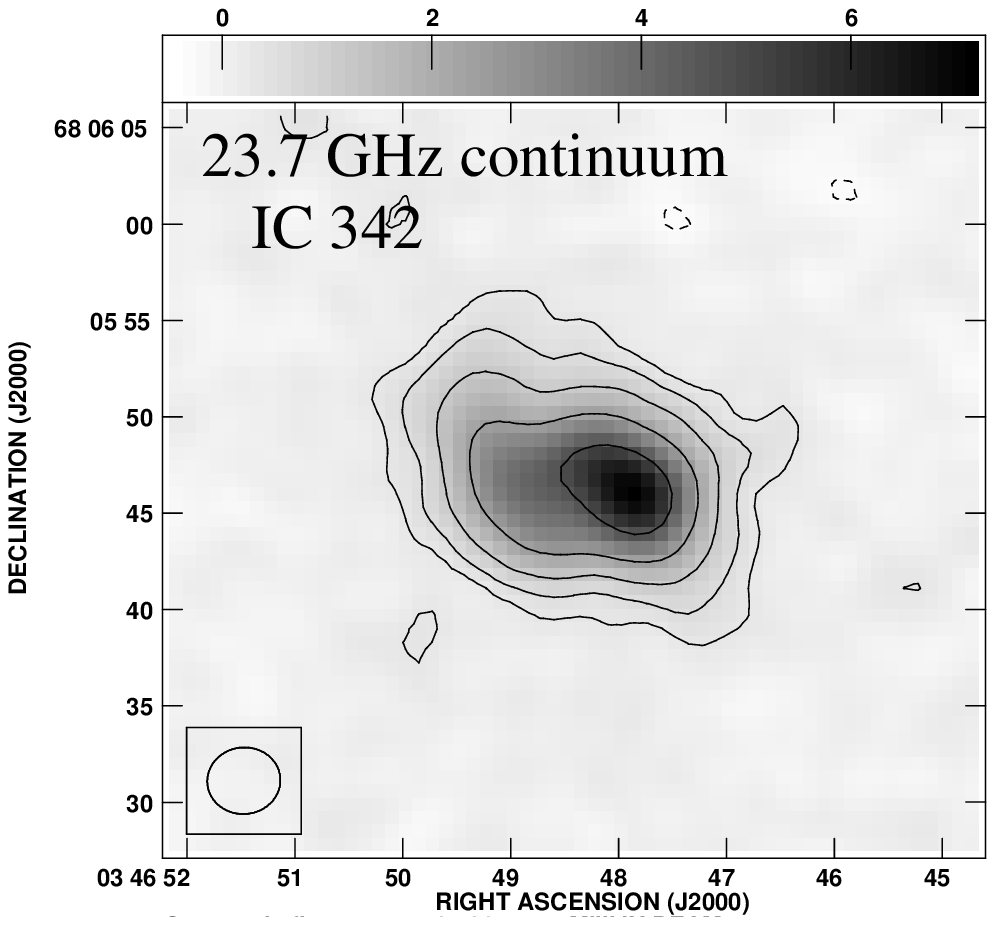,width=5cm}&
\epsfig{file=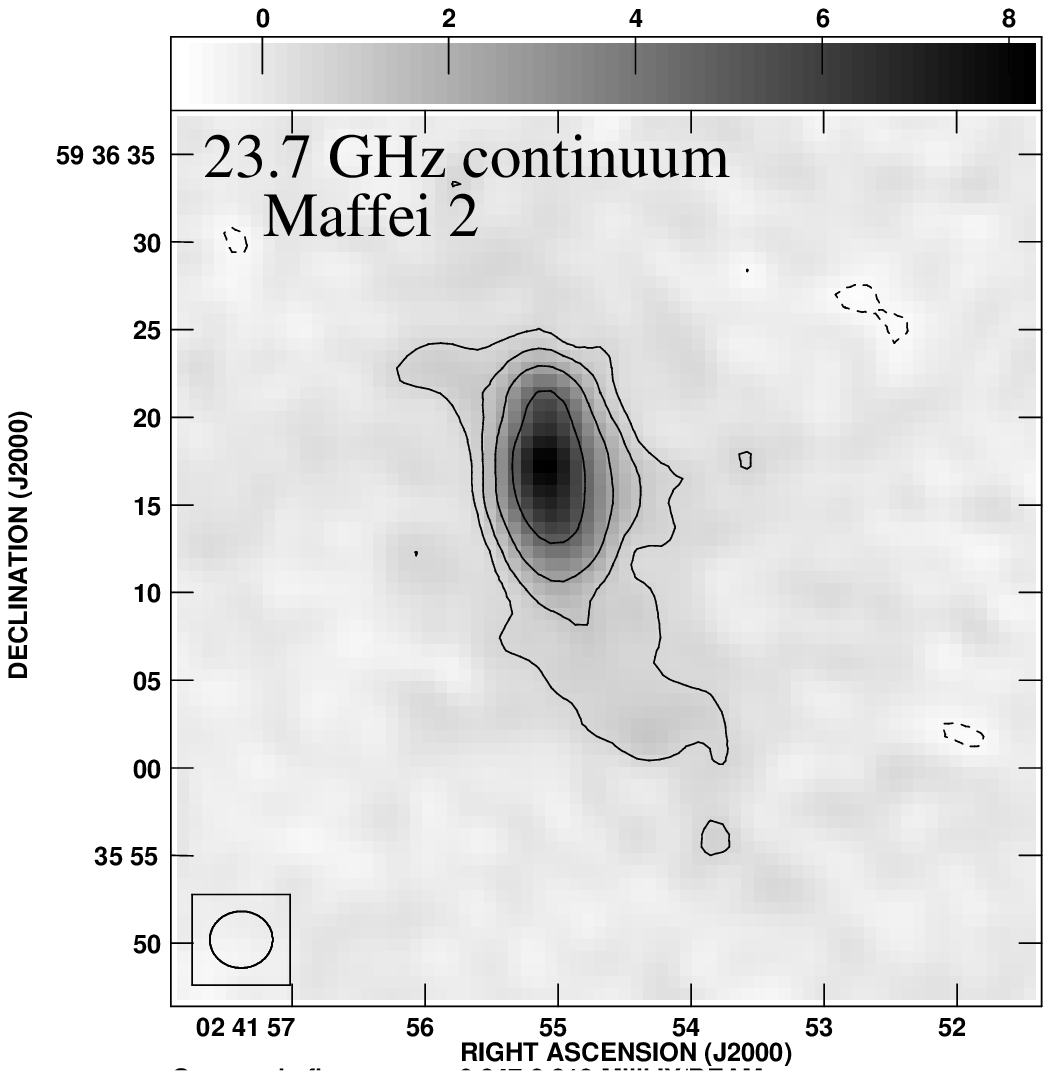,width=5cm}&
\epsfig{file=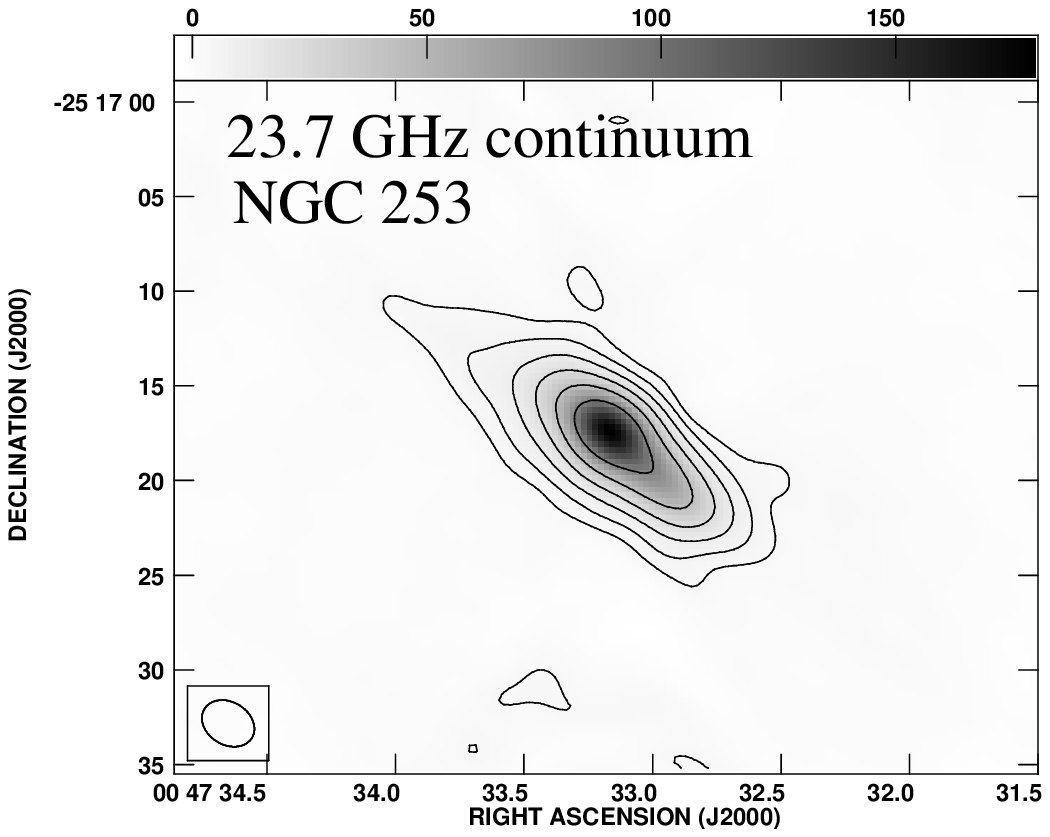,width=5cm}\\
 \end{tabular}
 \caption{Continuum maps at 23.7 GHz for IC\,342 (left), Maffei\,2 (middle),
 and NGC\,253 (right). The contour levels are,
$-0.3$, 0.3, 0.6, 1.2, 2.4, 4.8  mJy~beam$^{-1}$ for IC\,342,
$-0.6$, 0.6, 1.2, 2.4, 4.8 mJy~beam$^{-1}$ for Maffei\,2, and
$-2.7$, 2.7, 5.4, 10.8, 21.6, 43.2, 86.4  mJy~beam$^{-1}$ for
NGC\,253. The beams are shown at the bottom left corner of each
map.}\label{continuum}
\end{figure*}

\begin{figure*}[p]
 \begin{tabular}{c c}
  \epsfig{file=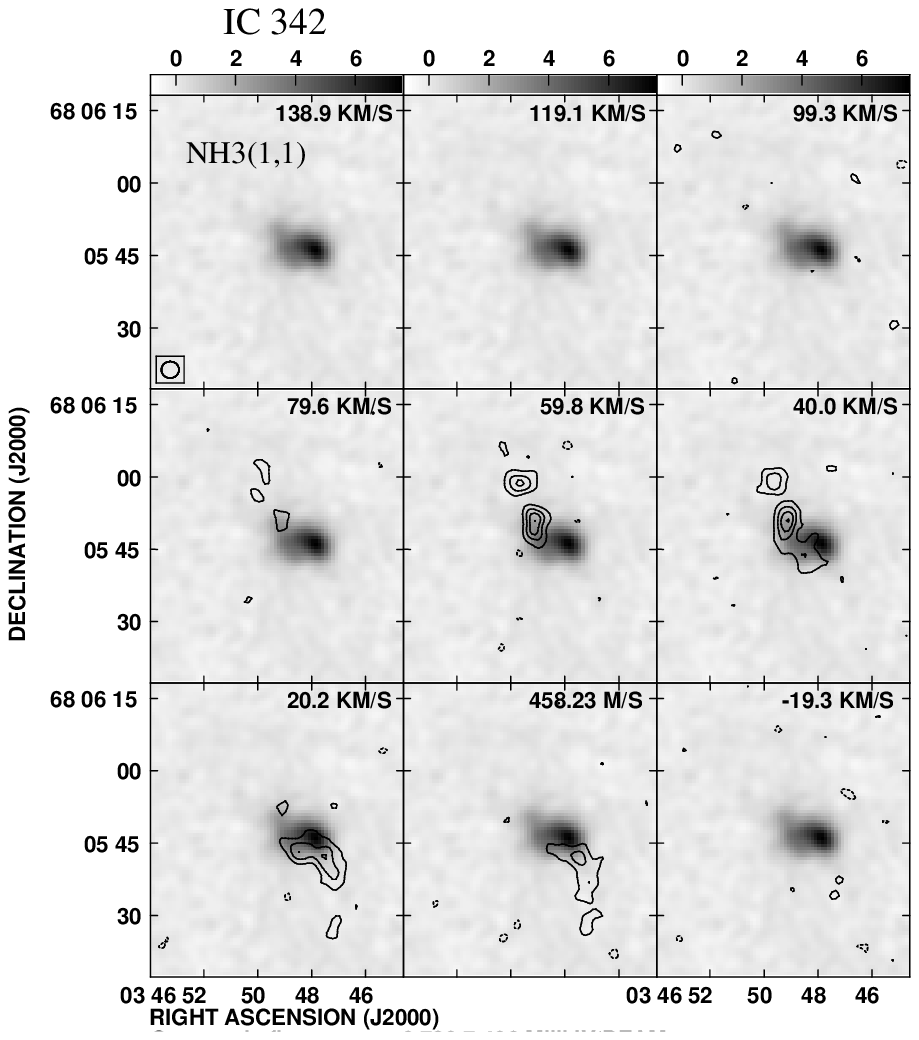,width=8cm}&
  \epsfig{file=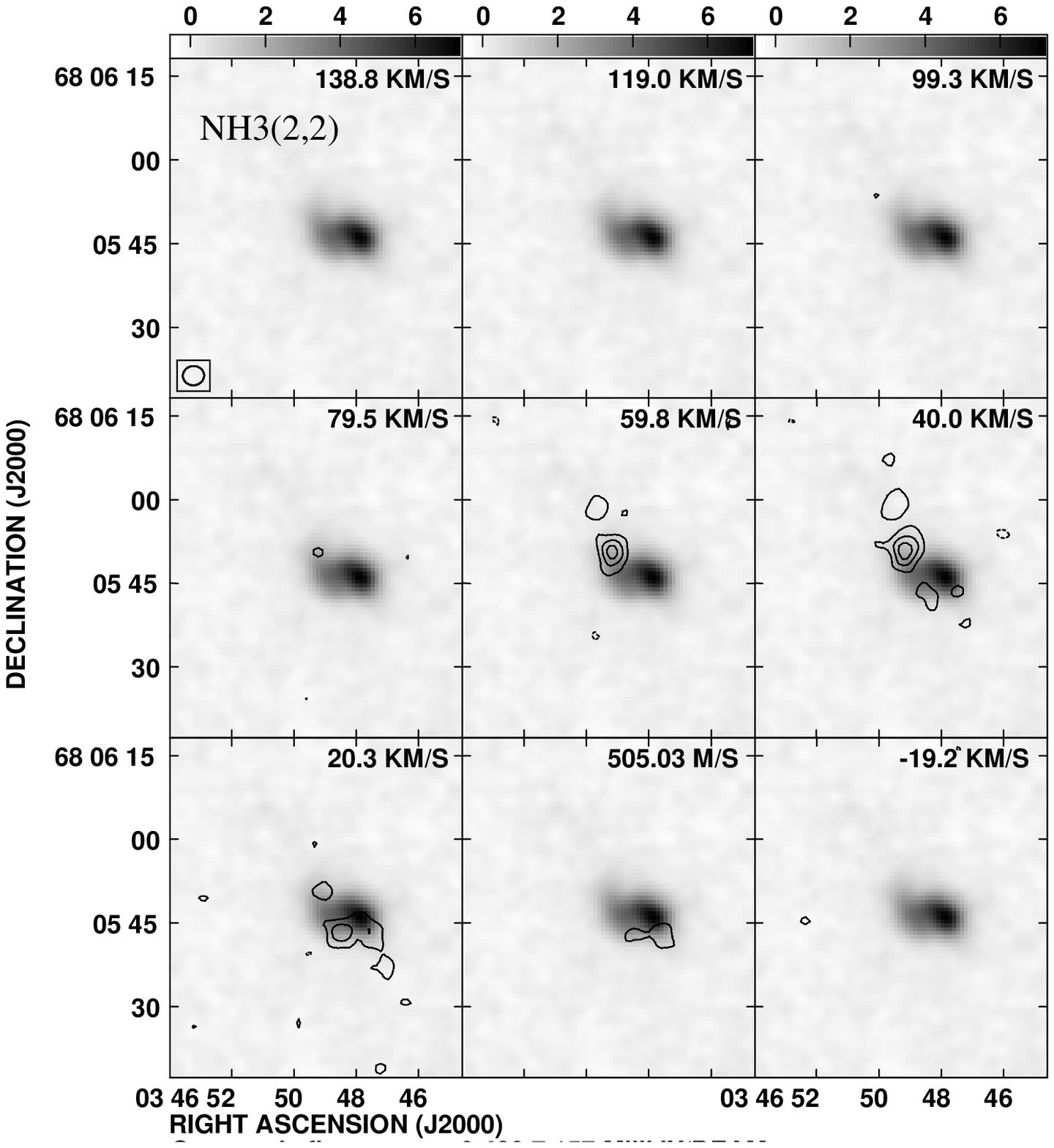,width=8cm}\\
 \end{tabular}
 \caption{Channel maps of the NH$_3$(1,1) and NH$_3$(2,2) transition
  emission from IC\,342.
The contour levels are $-$0.9, 0.9, 1.8, 2.7, and 3.6
mJy~beam$^{-1}$ and $-$0.75, 0.75, 1.5, 2.25, and 3 mJy~beam$^{-1}$,
respectively. The synthesized beams are shown in the lower left
corner of the first panel for each transition. The gray scale images
  represent the continuum emission at 23.7\,GHz.}
\label{ic342-ch}
\end{figure*}

\begin{figure*}[p]
 \begin{tabular}{c c}
  \epsfig{file=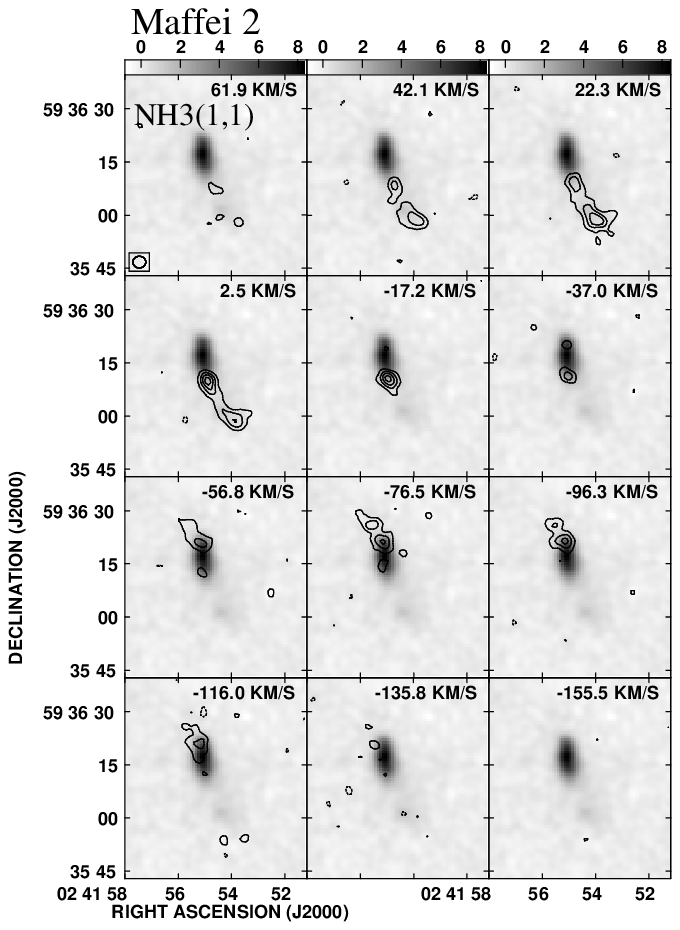,width=7cm}&
  \epsfig{file=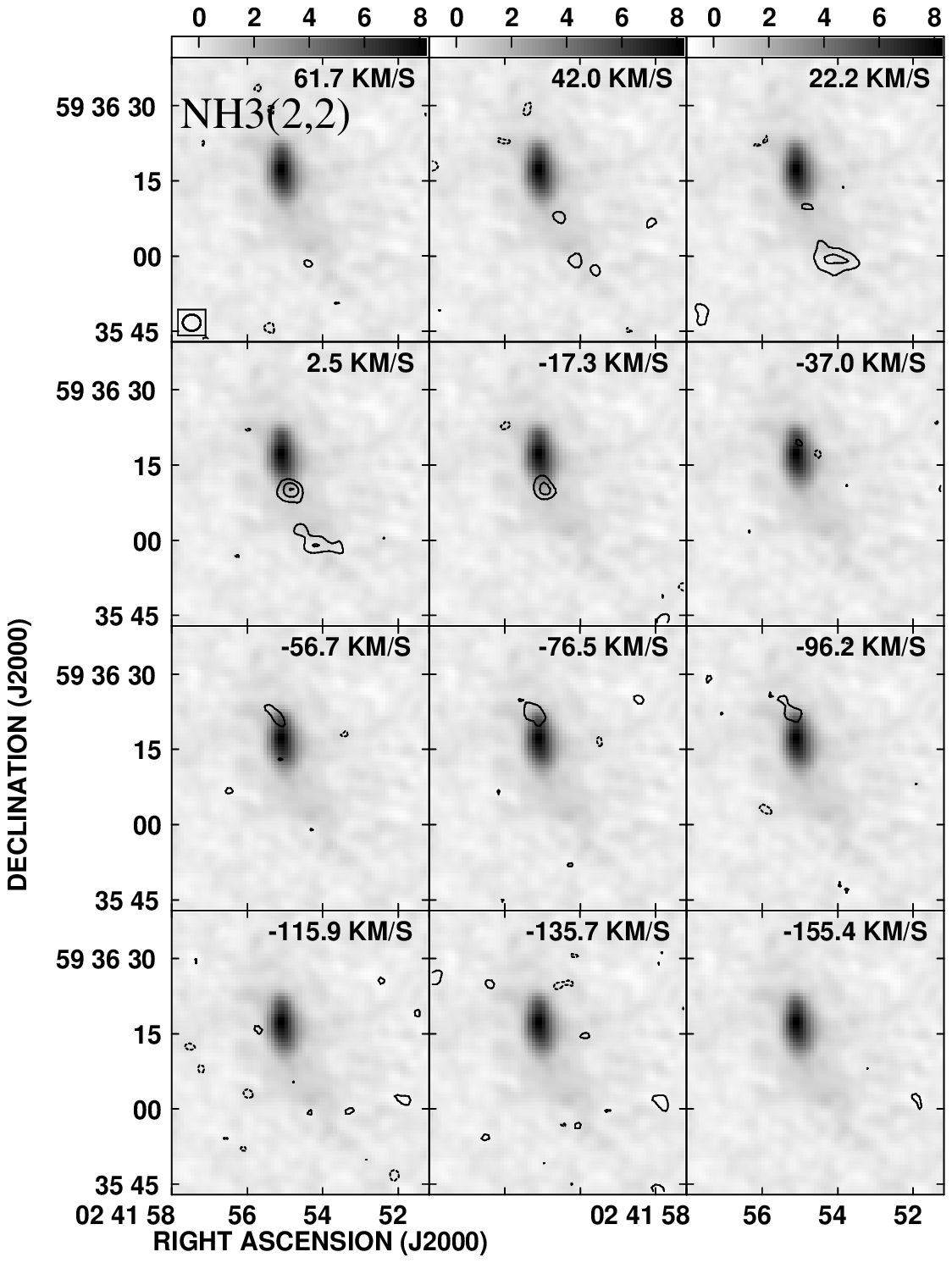,width=7cm}\\
 \end{tabular}
 \caption{Channel maps of the NH$_3$(1,1) and NH$_3$(2,2) emission from
Maffei\,2. The contour levels are $-$0.78, 0.78, 1.56, 2.34, and
3.12  mJy~beam$^{-1}$ and $-$1.2, 1.2, 2.4, 3.6, and 4.8
mJy~beam$^{-1}$, respectively.  The gray scale images show the
continuum emission at 23.7 GHz. Beams are
shown in the left upper panel for each map set.} \label{maffei2-ch}
\end{figure*}

\begin{figure*}[p]
\includegraphics[width=8cm]{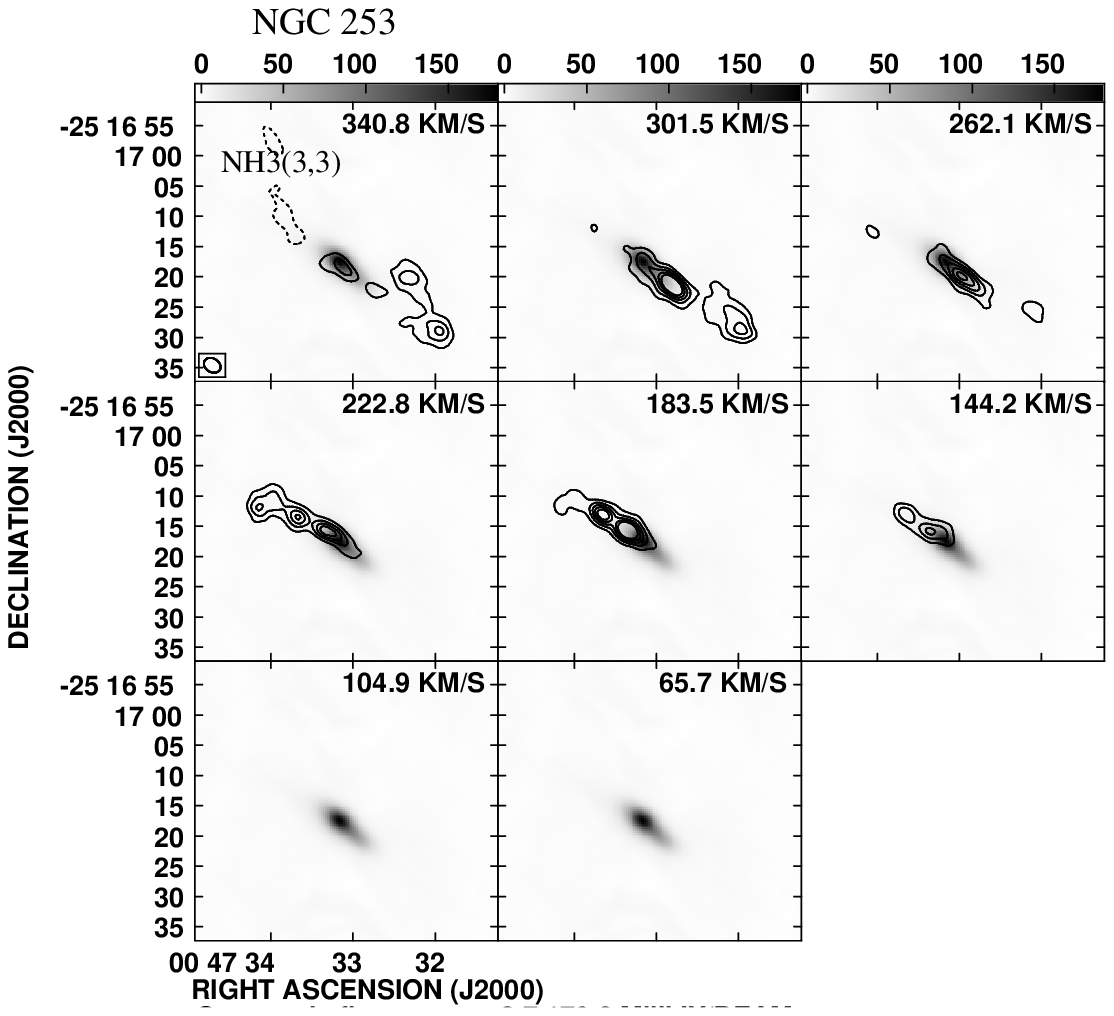}
\caption{Channel maps of the NH$_3$(3,3) transition from NGC\,253. The
contour levels are $-$1.8, 1.8, 3.6, 5.4, 7.2, and 9.0
mJy~beam$^{-1}$. The synthesized beam is shown in the left lower
corner in the first panel. The gray scale images represent the continuum
emission at 23.7 GHz.} \label{ngc253-33ch}
\end{figure*}


\begin{thebibliography}{}


\bibitem[2011]{aladro2011} Aladro, R., Mart\'in-Pintado, J., Mart\'in,
S., Mauersberger, R., \& Bayet, E. 2011, A\&A, 525, 89

\bibitem[2007]{beuther2007} Beuther, H., Walsh, A. J., Thorwirth, S.,
Zhang, Q., Hunter, T. R., Megeath, S. T., \& Menten, K. M. 2007, A\&A,
466, 989


\bibitem[2004]{brinchmann04} Brinchmann, C.S., White, S.D.M.,
et al. 2004, MNRAS 351, 1151


\bibitem[2009]{brunthaler2009} Brunthaler, A., 	Castangia, P., Tarchi,
A., Henkel, C., Reid, M. J., Falcke, H., \& Menten, K. M. 2009, A\&A,
497, 103

\bibitem[1983]{buta83} Buta, R.~J., \& McCall,
M.~L.\ 1983, \mnras, 205, 131



\bibitem[2010]{cost2010} Costagliola, F. \& Aalto, S. 2010, A\&A, 515,
71

\bibitem[2005]{combes2005} Combes, F. 2005, The evolution of
starbursts: The 331st Wilhelm and Else Heraeus Seminar. AIP
Conf. Proc. 783, 43





\bibitem[1988]{danby88} Danby, G., Flower, D. R., Valiron, P.,
Schilke, P., \& Walmsley, C. M. 1988, MNRAS, 235, 229

\bibitem[1992]{downes92} Downes, D., Radford, S.J.E., Guilloteau,
S., Gu\'elin, M., Greve, A., \& Morris, D., 1992, A\&A, 262, 424

\bibitem[2007]{Fingerhut2007} Fingerhut, R.L., Lee, H., McCall, M.L.,
\& Richer, M.G., 2007, ApJ 655, 814

\bibitem[1995]{flower95} Flower, D. R., Pineau des For\^ets, G., \&
Walmsley, C. M. 1995, A\&A, 294, 815

\bibitem[2001]{garciaburillo01}
Garc{\'{\i}}a-Burillo, S., Mart{\'{\i}}n-Pintado, J., Fuente, A.,
\& Neri, R.\ 2001, \apjl, 563, L27


\bibitem[1981]{geldzahler} Geldzahler, B.J. \& Witzel, A. 1981, AJ,
86, 1306


\bibitem[1989]{guesten89} G\"usten, R. 1989, in The Center of the Galaxy, ed: M. Morris, Kluwer, Dordrecht, p. 89




\bibitem[1995]{hartquist95} Hartquist, T.~W.,
Menten, K.~M., Lepp, S., \& Dalgarno, A.\ 1995, MNRAS, 272, 184

\bibitem[1987]{henkel87} Henkel, C., Jacq, T., Mauersberger, R., Menten, K. M., Steppe, H. 1987, A\&A, 188, L1


\bibitem[2000]{hen00} Henkel, C., Mauersberger, R., Peck, A.~B., Falcke, H., \& Hagiwara, Y. 2000, A\&A, 361, L45

\bibitem[2005]{henkel05} Henkel, C., Jethava, N., Kraus, A., et al. 2005, A\&A, 440, 893

\bibitem[2008]{henkel08} Henkel, C., Braatz, J. A., Menten, K. M., \& Ott, J. 2008, A\&A 485, 451

\bibitem[2002]{herrnstein2002} Herrnstein, R.M., Ho, P.T.P., 2002, ApJ 579, L83

\bibitem[1990]{ho90} Ho, P.~T.~P., Beck, S.~C., \& Turner, J.~L.\ 1990, \apj, 349, 57

\bibitem[1990a]{ho90a} Ho, P.~T.~P., Martin, R.~N., Turner, J.~L., Jackson, J.~M. 1990, ApJ, 355, L19

\bibitem[1983]{ho83} Ho, P.T.P. \& Townes, C.H. 1983, ARA\&A, 21, 239


\bibitem[1997]{hue97} H\"uttemeister, S., Mauersberger, R., Wilson, T.L. 1997, A\&A, 326, 59

\bibitem[1993]{huettemeister1993} H\"uttemeister, S., Wilson, T.~L., Bania, T.~M., \& Martin-Pintado, J.\ 1993, A\&A, 280, 255


\bibitem[1995]{hue95} H\"uttemeister, S., Wilson, T.L., Mauersberger, R., et al. 1995, A\&A, 294, 667


\bibitem[2007]{knudsen07} Knudsen, K.K., Walter F., Weiss, A., Bolatto, A., Riechers, D.A., \& Menten, K. 2007, ApJ, 666, 156

\bibitem[2000]{leteuff2000} Le Teuff, Y.~H., Millar, T.~J., \& Markwick, A.~J.\ 2000, A\&AS, 146, 157

\bibitem[2004]{lovas04} Lovas, F.J., 2004, J. Phys. Ref. Data, 33, 177


\bibitem[1986]{martin86} Martin, R.~N., \& Ho, P.~T.~P. 1986, ApJ, 308, L7

\bibitem[2006a]{martin06a} Mart\'in, S., Mart\'in-Pintado, J., \& Mauersberger, R. 2006, A\&A 450, L13

\bibitem[2006b]{martin06b} Mart\'in, S., Mauersberger, R., Mart\'in-Pintado, J., Henkel, C., Garc\'ia-Burillo, S. 2006, ApJS, 164, 450

\bibitem[1989]{mau89} Mauersberger, R. \& Henkel, C. 1989, A\&A, 223, 79

\bibitem[1991a]{mau91a} Mauersberger, R. \& Henkel, C. 1991, A\&A, 245, 457




\bibitem[1986]{mau86} Mauersberger, R., Wilson, T.~L.,\& Henkel, C. 1986, A\&A, 160, L13







\bibitem[1995]{mau95} Mauersberger, R., Henkel, C., \& Chin,
Y.-N. 1995, A\&A, 294, 23




\bibitem[2003]{mau03} Mauersberger, R., Henkel, C., Wei\ss, A., Peck, A.~B., \& Hagiwara, Y. 2003, A\&A, 403, 561

\bibitem[1989]{mccall89} McCall, M.L. 1989, AJ, 97, 1341

\bibitem[2002]{mcgary02} McGary, R.S., \& Ho, P.T.P. 2002, ApJ, 577,757

\bibitem[2000]{meier00} Meier, D.S., \& Turner, J.L. 2000, ApJ, 531, 200

\bibitem[2001]{meier01} Meier, D.S., \& Turner, J.L. 2001, ApJ, 551, 687

\bibitem[2005]{meier05} Meier, D.~S., \& Turner, J.~L.\ 2005, \apj, 618, 259

\bibitem[2008]{meier08} Meier, D.~S., Turner, J.~L., \& Hurt, R.~L.\ 2008, \apj, 675, 281


\bibitem[2006]{montero2006} Montero-Casta{\~n}o, M., Herrnstein, R.~M., \& Ho, P.~T.~P.\ 2006,
\apj, 646, 919

\bibitem[1991]{nguyen91} Nguyen-Q-Rieu, Henkel, C., Jackson, J.M., \& Mauersberger, R. 1991, A\&A, 241, L33

\bibitem[2005]{oka2005} Oka, T., Geballe, T.~R., Goto, M., Usuda, T., \& McCall, B.~J.\ 2005, \apj, 632, 882



\bibitem[2005]{ott05b} Ott, J., Wei\ss, A., Henkel, C., \& Walter, F. 2005, ApJ, 629, 767


\bibitem[2005]{Rekola2005} Rekola, R., Richer, M. G., McCall, M. L.,
Valtonen, M. J., Kotilainen, J. K., \& Flynn, C. 2005, MNRAS, 361, 330

\bibitem[1983]{rickard83} Rickard, L.~J., \& Harvey, P.~M.\ 1983, \apjl, 268, L7

\bibitem[2010]{Riquelme2010} Riquelme, D., Bronfman, L., Mauersberger,
R., May, J., Wilson, T.L., 2010, A\&A 523, 45


\bibitem[1996]{rohlfs96} Rohlfs, K. \& Wilson, T.L. 1996, Tools of Radioastronomy, Springer Verlag


\bibitem[2002]{saha2002} Saha, A., Claver, J, \& Hoessel, J.G.  2002, AJ, 124, 838


\bibitem[1999]{sakamoto99} Sakamoto, K., Okumura, S.~K., Ishizuki, S., \& Scoville, N.~Z.\ 1999, ApJ, 525, 691

\bibitem[2006]{sakamoto06} Sakamoto, K., Ho, P.T.P., Daisuke, I., et al.\ 2006, \apj, 636, 685

\bibitem[2011]{Sakamoto2011} Sakamoto, K., Mao, R., Masushita, S.,
Peck, A.B., Sawada, T., Wiedner, M.W., 2011, ApJ, 735, 19

\bibitem[1974]{sandage74} Sandage, A., \& Tammann, G. A. 1974, ApJ, 194, 559




\bibitem[1995]{sternberg95} Sternberg, A., \& Dalgarno, A. 1995, ApJS, 99, 565

\bibitem[1983]{Suto1983} Suto, M. \& Lee, L. C., 1983, J.~Chem.~Phys.,
78, 4515


\bibitem[2002]{takano02} Takano, S., Nakai, N., Kawaguchi, K. 2002, PASJ, 54,
195

\bibitem[2000]{takano00} Takano, S., Nakai, N., Kawaguchi, K., \& Takano, T. 2000, PASJ, 52, L67

\bibitem[2005a]{Takano05a} Takano, S., Hofner, P., Winnewisser, G.,
Nakai, N., \& Kawaguchi, K. 2005, PASJ, 57, 549

\bibitem[2005b]{takano2005b} Takano, S., Nakanishi, K., Nakai, N., \& Takano, T.\ 2005, PASJ, 57, L29

\bibitem[2002]{tarchi02} Tarchi, A., Henkel, C., Peck, A.~B., \& Menten, K.~M.\ 2002, A\&A, 385, 1049

\bibitem[2010]{tikhonov2010} Tikhonov, N. A. \& Galazutdinova,
O. A. 2010, AstL, 36, 167


\bibitem[1983]{turner83} Turner, J. L., \& Ho P.~T.~P. 1983, ApJ, 268, L79


\bibitem[1994]{turner94} Turner, J. L., \& Ho P.~T.~P. 1994, ApJ, 421, 122

\bibitem[1997]{ulvestad97} Ulvestad, J.S. \& Antonucci, R.R.J. 1997,
ApJ, 488, 621

\bibitem[2004]{usero04} Usero, A., Garc{\'{\i}}a-Burillo, S., Fuente, A., Mart{\'{\i}}n-Pintado, J., \& Rodr{\'{\i}}guez-Fern{\'a}ndez, N.~J.\ 2004, \aap, 419, 897

\bibitem[2006]{usero06} Usero, A., Garc\'ia-Burillo, S., Mart\'in-Pintado, J., Fuente, A., \& Neri, R., 2006, A\&A 448, 457


\bibitem[1983]{walmsley83} Walmsley, C.~M., \& Ungerechts, H. 1983, A\&A, 122, 164

\bibitem[2001a]{weiss01a} Wei{\ss}, A., Neininger, N., Henkel, C., Stutzki, J., \& Klein, U. 2001, ApJ, 554, L143


\bibitem[2007]{weiss2007} Wei{\ss}, A., Downes, D., Neri, R., Walter,
F., Henkel, C., Wilner, D. J., Wagg, J., \& Wiklind, T. 2007, A\&A,
467, 955

\bibitem[1990]{wilson90} Wilson, T.~L., \& Mauersberger, R.\ 1990, \aap, 239, 305

\bibitem[2006]{wilson2006} Wilson, T. L., Henkel, C., \&
H\"uttemeister, S. 2006, A\&A, 460, 533

\bibitem[1989]{ziurys89} Ziurys, L.~M., Friberg, P., \& Irvine, W.~M.\ 1989, \apj, 343, 201


\end{thebibliography}
\end{document}